\documentclass{PoS}

\usepackage{amsmath}
\usepackage{subfigure}

\title{Review of minimum-bias jet systematics at RHIC}

\ShortTitle{"Review of minimum-bias jet systematics at RHIC"}

\author{\speaker{Duncan Prindle} for the STAR Collaboration%
         \\
        University of Washington\\
        E-mail: \email{prindle@npl.washington.edu}}

\abstract{
    Jets are studied in A-A collisions at RHIC and LHC with the goal to understand
    how they are affected by the medium and how they affect the medium.
    It is widely believed that hard-scattered partons lose energy when propagating
    through a medium before hadronizing.
    Partons losing enough energy may not even make it out of the medium as identifiable jets
    (although the momentum will be shared among whatever particles are emitted).
    ``Full'' jet reconstruction attempts to determine the partonic energy loss as well
    as possible changes in jet shape.
    Heavy ion collisions typically produce many unrelated particles within the
    jet ``cone,'' and subtraction of this background introduces significant uncertainties.
    A variety of techniques using high-$p_t$ particles, assumed to be leading particles from
    jet fragmentation, look for disappearance of jets and attenuation of jets relative to the
    reaction plane, as well as medium modifications such as Mach cones.
    Those techniques have considerable uncertainty due to subtraction of $\rm v_2$.
    In this paper we discuss minimum-bias jets observed at RHIC using two-particle correlations.
    We find that jets produced in p-p collisions have interesting properties.
    Peripheral A-A collisions look like p-p collision.
    As we select more central collisions the number of jets increases following binary collision scaling
    until at a system-dependent centrality the number of particles associated with jets increases substantially
    above this scaling.
    Near this transition centrality the jet aspect ratio---elongated transverse to the beam direction
    for low-energy jets produced in p-p collisions---becomes highly elongated along the
    beam direction in A-A collisions.
}

\FullConference{Workshop on Critical Examination of RHIC Paradigms\\
		 April 14, 2010\\
		 Austin, Texas, U.S.A.}

\begin{document}

\section{Introduction}
Our studies of two-particle correlations began in the late 1990's with the desire to understand
heavy-ion collisions in an unbiased way~\cite{TrainorCentralLimit}.
At that time it was expected that central A-A collisions would be nearly thermalized with jets
quenched and, if not completely absorbed, difficult to observe.
Temperatures inferred from $p_t$ spectra could fluctuate event to event or from place to place within
an event, the amplitude and volume of $p_t$ fluctuations depending on the nearness of the event to the
quark-gluon plasma phase boundary.
We wanted to observe these fluctuations in an unambiguous way.

A number of techniques were considered, such as factorial moments~\cite{dremin-kittel}~\cite{bialas},
wavelets~\cite{dremin}
and entropy transport~\cite{TrainorReid}.
But those measurements are logarithmic in the scale, and we soon realized that the range of scales available
to a detector like STAR was at best two orders of magnitude, making a measure linear in scale preferable.
We eventually settled on measures related to variance differences and learned how to convert
between correlations and fluctuations.~\cite{fluctInversion}
These correlation measures can be related to Pearson's correlation coefficient, making them
more easily interpretable than fluctuations which are integrals over correlations and thus include
correlation structure from many scales.

There were some concerns about whether such a general approach could work, partly fueled by 
previous difficulty interpreting factorial moments~\cite{bialasQM}~\cite{seybothHBT}.
Also, it is easy for a detector artifact to produce a signal which may confuse the physics interpretation.
With a variance-based measure it is possible to cancel detector artifacts in a natural
way using mixed events.
It was also believed that a generic two-particle correlation analysis wouldn't be sensitive
enough to see correlations such as flow.
Monte Carlo simulations of temperature fluctuations within nearly-thermalized events were performed and it was
shown that our techniques could measure the fluctuations that were expected at that time.
In fact we could distinguish the case where events were thermalized with each event having a
different temperature from the case where events were nearly thermalized but having a temperature that
varied with position~\cite{mevsim}.
When we analyzed collision data we saw considerable correlation structure, but no evidence for
temperature fluctuations~\cite{aya}.

There are several structures apparent in two-particle correlations.
These are typically different enough in angular size that we can disentangle them.
There is a very sharp $e^+e^-$ peak due to $\gamma$-induced pair production.
The observed HBT peak is broader than the $e^+e^-$ peak but narrower than
the same-side (SS) 2D peak due to minijets.
When they are similar sizes we can use the fact that HBT is only observed for like-sign (LS)
pairs while minijets have a strong unlike-sign (US) component.
Back-to-back scattered partons are observed as an away-side (AS) ridge (as well as both partons
contributing to the SS 2D peak) and this ridge is approximately independent of
$\eta_\Delta = \eta_1 - \eta_2$ within the STAR TPC acceptance.
For high $p_t$ particles we expect the AS ridge to be a Gaussian centered at $\pi$ on $\phi_\Delta = \phi_1 - \phi_2$,
but when that Gaussian is broad enough in $\phi_\Delta$ (as it is for most of our $p_t$ range)
it has a $\cos(\phi_\Delta - \pi)$ shape.
We refer to $\cos(\phi_\Delta - \pi)$ as an azimuth dipole.
There is also a quadrupole observed, $\cos(2\phi_\Delta)$.
The dipole and quadrupole are orthogonal to each other on $\phi_\Delta$ and distinguished from HBT
and the SS 2D Gaussian minijet peak by their long ranges in $\eta_\Delta$.

In the rest of this paper we emphasize the jet-like components of two-particle correlations.
We start with a discussion of two-particle correlation measurement techniques, noting 
the equivalence of fluctuations and correlations, and discuss how to interpret
the multi-dimensional two-particle correlation space.
Then we examine p-p collisions, the reference system
for A-A collisions, and find interesting two-particle correlation structure.
We next see how some of this structure is modified in A-A collisions.
Finally we make some comments about how $p_t$ correlations complement
number correlations.

\section{Review of Methods}
Jets can be defined by a jet-finding algorithm which groups all particles
that come from the fragmentation of a hard-scattered parton~\cite{DokJetAlgo}\cite{EllisJetAlgo}\cite{FastJet}.
There are a few commonly used algorithms.
It is an active area of research
to determine which ones work best in a heavy ion collision environment.
How well these algorithms work depends on the jet shape, which could depend on a jet-medium
interaction as well as the fluctuating background of hadrons unrelated to the particular scattered parton.
An alternative to explicit jet reconstruction is to study two-particle correlations.
This approach makes no assumption about jet shapes and allows one to study lower-energy
partons than explicit jet reconstruction algorithms permit.
Indeed, below a few GeV, partons will only be able to fragment into two hadrons,
and this is difficult to observe with jet reconstruction algorithms
in heavy ion environments.
Another advantage of two-particle correlations is that they
allow study of inter-jet correlations as well as intra-jet correlations.

The most common measure of the dependence between two statistical quantities is
Pearson's correlation coefficient~\cite{Pearsons}, the covariance divided by the geometric mean of variances,
which is bounded by -1 and 1.
We are interested in the structure of two-particle momentum-space correlations.
We write the covariance $\Delta\rho$ as a difference between
an object and a reference,
\begin{align*}
    \Delta\rho (\vec{p_1},\vec{p_2}) &= \rho_{sib} (\vec{p_1},\vec{p_2}) - \rho_{ref} (\vec{p_1},\vec{p_2})
    \\                               &= \rho_{(2)} (\vec{p_1},\vec{p_2}) - \rho_{(1)} (\vec{p_1}) \rho_{(1)} (\vec{p_2})
\end{align*}
where $\rho_{(2)}$ is a two-particle density and $\rho_{(1)}$ is a one-particle density.
In practice we bin quantities, storing them in histograms.
For the case of number correlations the bin $(a, b)$, representing correlations between particles
at positions a and b, can be written as
\begin{equation*}
\epsilon_a\epsilon_b\Delta\rho(n) = \left< n_a n_b \right> - \left< n_a \right> \left< n_b \right>
                                  = \overline{(n - \bar{n})_a (n - \bar{n})_b}
\end{equation*}
where $\epsilon$ are the bin widths.
For Pearson's correlation coefficient we divide by the geometric mean of the variances,
$\Delta\rho/\sqrt{\sigma_{n_a}^2 \sigma_{n_b}^2} \approx \Delta\rho/\sqrt{\bar{n}_a\bar{n}_b} = \Delta\rho/\sqrt{\rho_{ref}}$.
We have replaced the variances in the denominator with the Poisson expectations so that all the
physics is in the numerator.
In a detector one must deal with efficiencies and acceptances.
The quantity we actually use is thus
\begin{equation*}
    \sqrt{\rho_{ref}^{'}} \frac{\Delta\rho}{\rho_{ref}}.
\end{equation*}
We normally refer to this simply as $\Delta\rho/\sqrt{\rho_{ref}}$.
We define $\sqrt{\rho_{ref}^{'}} \equiv \frac{d^2N}{d\eta d\phi}$.
The ratio $\Delta\rho/\rho_{ref}$
has the virtue of canceling efficiencies and acceptances.
Besides being closely related to a standard correlation measure the quantity $\Delta\rho/\sqrt{\rho_{ref}}$
is the correlation strength per final-state particle.
If A-A collisions are simply superpositions of
N-N collisions it will be independent of centrality and collision system.

\begin{figure}
    \subfigure[$\Delta\sigma^2_{p_t:n}$ at STAR full scale]{\includegraphics [width=0.3\textwidth]{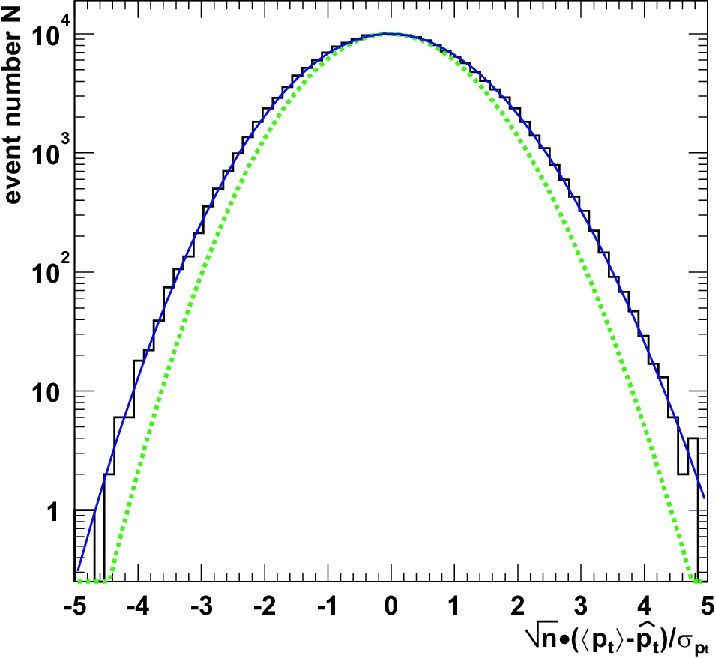}}
    \subfigure[scaled fluctuations $\Delta\sigma^2_{p_t:n}$]{\includegraphics [width=0.3\textwidth]{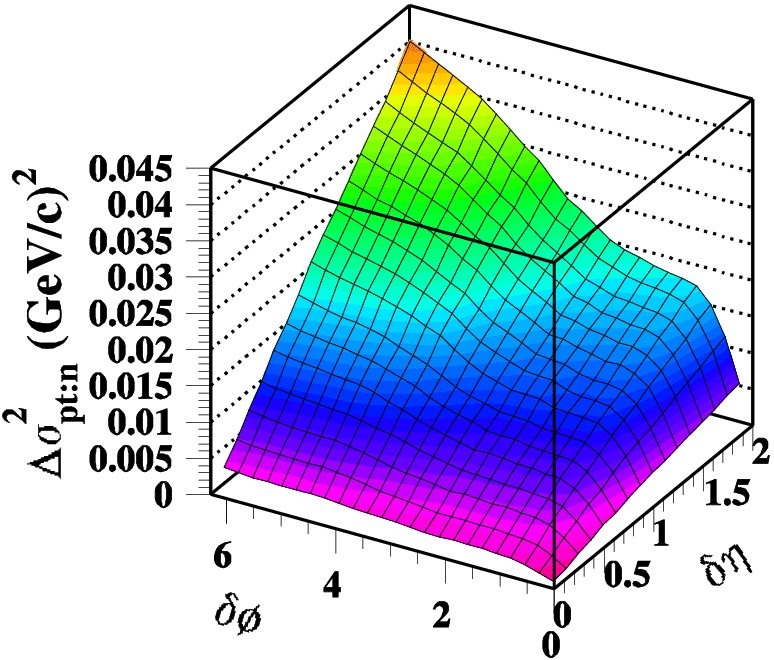}}
    \subfigure[$\Delta\rho/\sqrt{\rho_{ref}}$]{\includegraphics [width=0.3\textwidth]{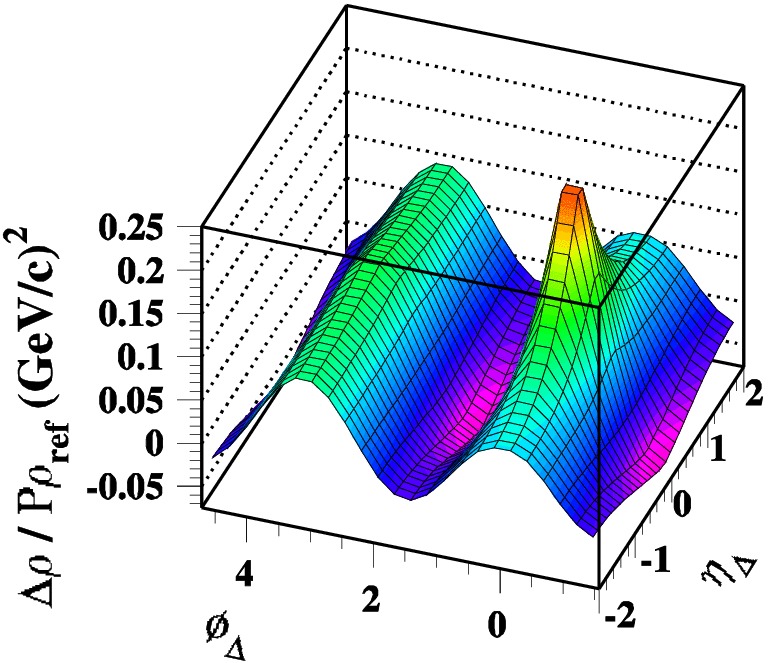}}
    \caption{Sequence going from measurement of fluctuations at STAR full scale to scaled
             fluctuations probing dependence on size and finally to correlations inferred from fluctuations.
             There is an interesting structure in the correlations.
    }
    \label{fig1}
\end{figure}

Correlations and fluctuations are closely related, fluctuations resulting from an
integral over the correlation structure.
For the angular space the integral equation relating the fluctuations and correlations is~\cite{fluctInversion}
\begin{equation*}
    \Delta\sigma^2_{p_t;n}(\delta\eta,\delta\phi) = 4\int_0^{\delta\eta}d\eta_\Delta\int_0^{\delta\phi}d\phi_\Delta K(\delta\eta,\delta\phi;\eta_\Delta,\phi_\Delta) \frac{\Delta\rho}{\sqrt{\rho_{ref}}}(\eta_\Delta,\phi_\Delta),
\end{equation*}
where $\Delta\sigma^2_{p_t;n}(\delta\eta,\delta\phi)$ is a variance difference measured in a ``detector'' of
size $(\delta\eta,\delta\phi)$, the kernel $K$ is an exactly known geometric factor and $\Delta\rho/\sqrt{\rho_{ref}}$
is a two-particle autocorrelation.
This form of integral equation is known as a Volterra equation.
Writing this equation in terms of binned quantities we can evaluate the kernel explicitly (Eq.~5 of~\cite{fluctCorr})
\begin{equation*}
    \Delta\sigma^2_{p_t;n}(\delta\eta,\delta\phi) = 4\sum_{k=1}^{k=m}\sum_{l=1}^{l=n}\left( 1-\frac{k-1/2}{m}\right) \left( 1-\frac{l-1/2}{n}\right) \frac{\Delta A}{\sqrt{A_{ref}}}(k\epsilon_\eta,l\epsilon_\phi).
\end{equation*}
Here we use $A$ instead of $\rho$ to indicate it is a binned quantity.
We see that in general fluctuations depend on the domain scale over which they are measured.
When we measure fluctuations at a particular scale we have an integral of the correlations up to that scale.
The difference in fluctuations between two scales is a measure of the integral of the correlations between
those scales.
We can measure the scale dependence of $\Delta\sigma^2_{p_t;n}$ on $(\delta\eta,\delta\phi)$ and invert
the integral equation to infer $\frac{\Delta\rho}{\sqrt{\rho_{ref}}}(\eta_\Delta,\phi_\Delta)$.
Inverting Volterra equations (to solve for correlations from fluctuations) is an example of an
ill-posed problem.
One must impose a regularization scheme to obtain a useful solution, analogous to applying a low-pass filter.
There is a well-developed mathematical framework for this~\cite{tychonoff} but
one always has to assess how much the regularization scheme affects the extracted correlations.
An example of inverting scaled fluctuations to get correlations is shown in Fig.~\ref{fig1}~\cite{ptFluct}\cite{fluctInvPaper}.
We note that one motivation for measuring mean-$p_t$ fluctuations was to determine event-to-event
temperature fluctuations.
The correlation structure we observe is inconsistent with temperature fluctuations, instead it
reveals jet correlations.

Fluctuations are less numerically expensive to compute than correlations but they are difficult to interpret.
We can invert the integral equation to get the more-easily interpretable correlation structure, but
the issue with regularization will always be a source of uncertainty.
There are also technical issues such as two-track resolution that are difficult to correct in a fluctuation measure.
Having established a direct connection~\cite{fluctInversion} we prefer to measure angular correlations directly.

Each particle has three momentum components, so the two-particle correlation space is
six dimensional.
In a detector with complete azimuthal coverage the absolute position of the pair
($\phi_\Sigma = \phi_1 + \phi_2$) does not matter.
Only the difference $\phi_\Delta$ matters.
Within the $\eta$ acceptance of the STAR TPC it is approximately true that pair correlations
are independent of $\eta_\Sigma \equiv \eta_1 + \eta_2$, depending only on difference $\eta_\Delta$.
Averaging over $\phi_\Sigma$ and $\eta_\Sigma$ reduces the dimensionality of the correlation space to four
with no loss of information.

\begin{figure}
    \subfigure[$(\eta_\Delta,\phi_\Delta)$ projection]{\includegraphics [width=0.245\textwidth]{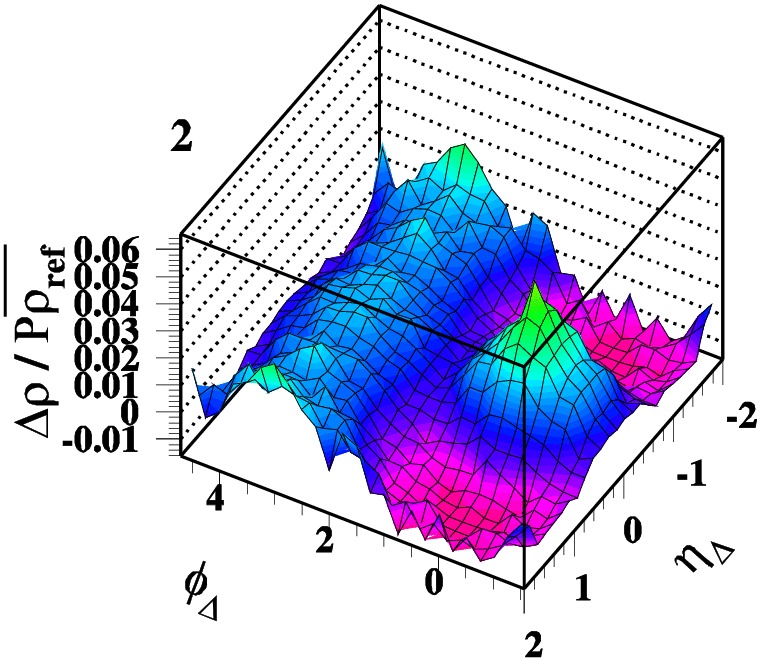}\label{fig2_a}}
    \subfigure[$(y_{t1},y_{t2})$ projection]{\includegraphics [width=0.245\textwidth]{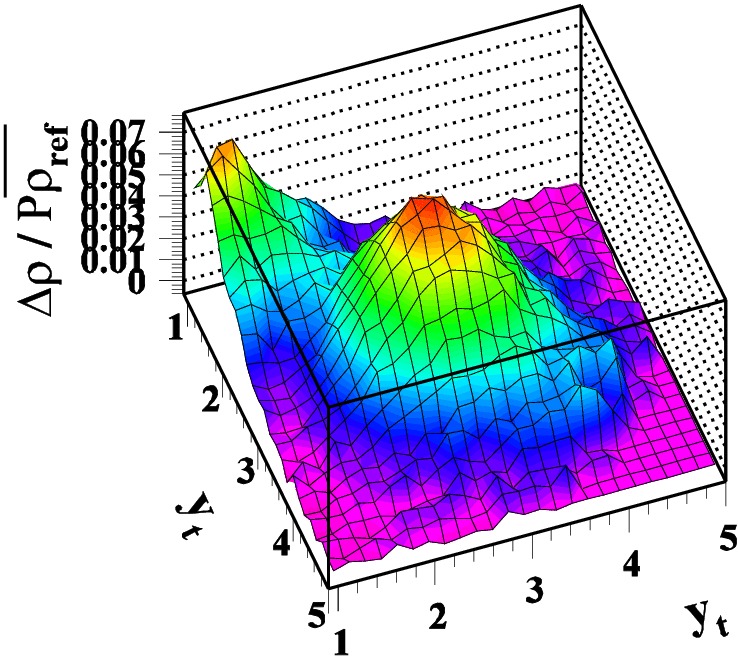}\label{fig2_b}}
    \subfigure[US, AS projection]{\includegraphics [width=0.245\textwidth]{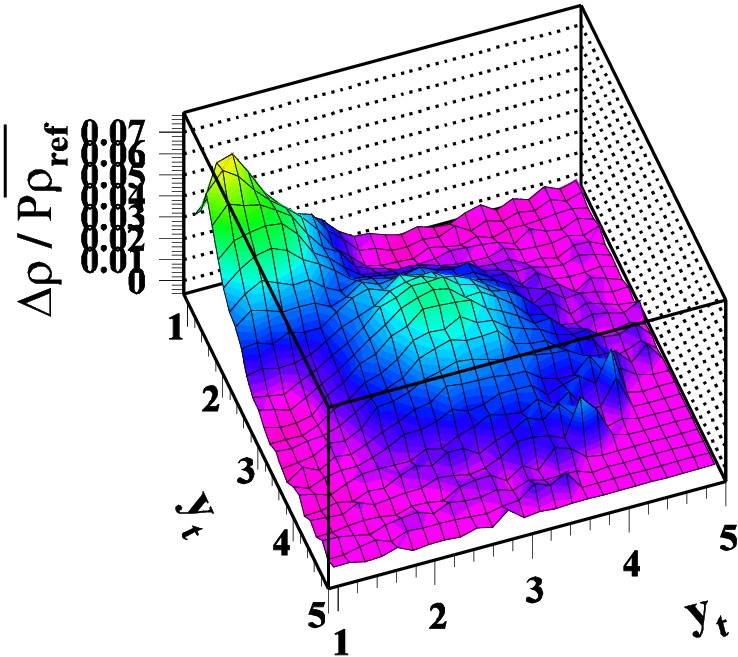}\label{fig2_c}}
    \subfigure[US, SS projection]{\includegraphics [width=0.245\textwidth]{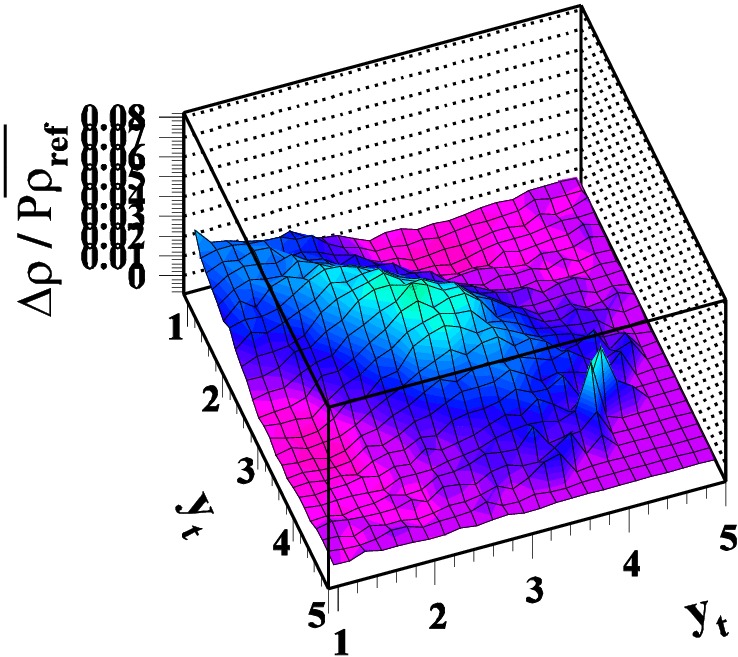}\label{fig2_d}}
    \caption{Projections of the six dimensional correlation space onto two dimensional subspaces for p-p collisions.
             Panel~\protect\ref{fig2_a} is the projection onto the angular subspace, $(\eta_\Delta,\phi_\Delta)$, with a ``hard'' $p_t$ cut imposed.
             We see an intra-jet correlation on the same-side (SS) and an inter-jet ridge on the
             away-side (AS); the ridge is due to the fact that the parton center of mass is not at rest
             in the laboratory frame.
             Panel~\protect\ref{fig2_b} is the full space projected onto the $(y_{t1},y_{t2})$ subspace.
             Panels~\protect\ref{fig2_c} and \protect\ref{fig2_d} are projections of the AS and SS correlations of unlike-sign (US) pairs
             onto the $(y_{t1},y_{t2})$ subspace.
    }
    \label{fig2}
\end{figure}

We are still left with a four-dimensional space, two angle differences and two momentum magnitudes.
It is common to use $p_t$ for the momentum magnitude, but since the yield falls steeply with $p_t$
most of the pairs occupy a small corner of the relevant $(p_{t1},p_{t2})$ space.
We use $y_t \equiv \ln\left[ \left( m_t+p_t\right) /m_\pi\right]$ for better visual access to the
low-$p_t$ correlation structure.
Four dimensions is impossible to visualize so we project onto the 2D angular subspace,
$(\eta_\Delta,\phi_\Delta)$, which has the form of a joint autocorrelation,
or the 2D $(y_{t1},y_{t2})$ subspace.
It is instructive to make cuts on the space that is projected out.
In Fig.~\ref{fig2} we show examples of projections onto the $(\eta_\Delta,\phi_\Delta)$ and $(y_{t1},y_{t2})$ subspaces.
For the $(y_{t1},y_{t2})$ subspace we show the projection over the full angular space as well
as for the AS and the SS.

\section{Two-particle correlations from proton-proton collisions}

In this section we examine the two-particle correlation structure of particles produced in
p-p collisions.
First we make a connection with a spectrum analysis in which events with different numbers of
observed charged particles ($n_{ch}$) were analyzed separately.
It was observed that one can define a ``soft'' component.
When the soft component is subtracted from each of the $n_{ch}$
spectra what remains has a shape and position independent of $n_{ch}$, but an amplitude proportional
to $n_{ch}$---the so-called ``hard'' component~\cite{ppTwoComponent}.
When plotted on $y_t$ the hard-component shape is Gaussian with a peak at $y_t = 2.7$ or $p_t = 1\rm\ GeV/c$.
This is the same position of the peak we find in the projection of the two-particle
correlations onto $(y_{t1},y_{t2})$ as shown in Fig.~\ref{fig2_b}.
In Fig.~\ref{fig4} we will see that for low-$y_{t\Sigma} \equiv y_{t1} + y_{t2}$ 
pairs the angular correlations
are dominated by soft physics while for high-$y_{t\Sigma}$ pairs the angular correlations are
dominated by hard scattering.

\begin{figure}
        \subfigure[SS, LS]{\includegraphics [width=0.245\textwidth]{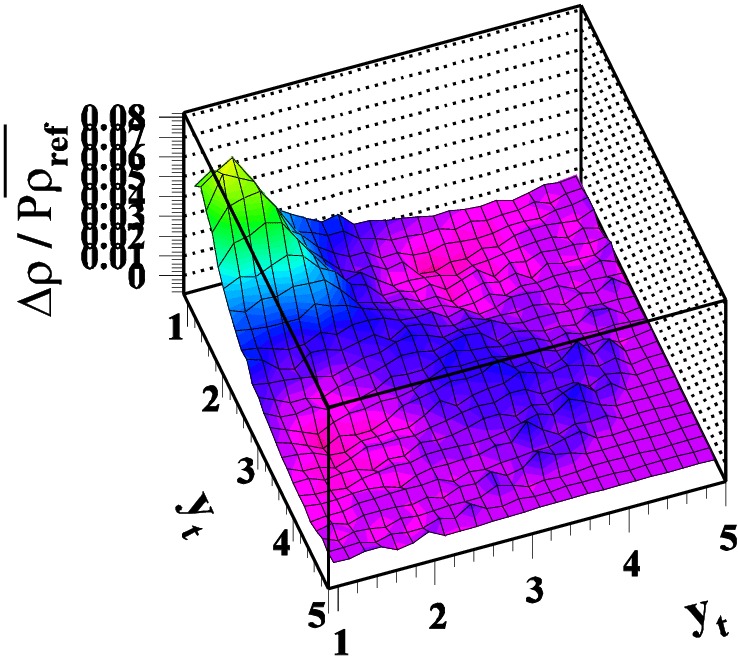}}
        \subfigure[SS, US]{\includegraphics [width=0.245\textwidth]{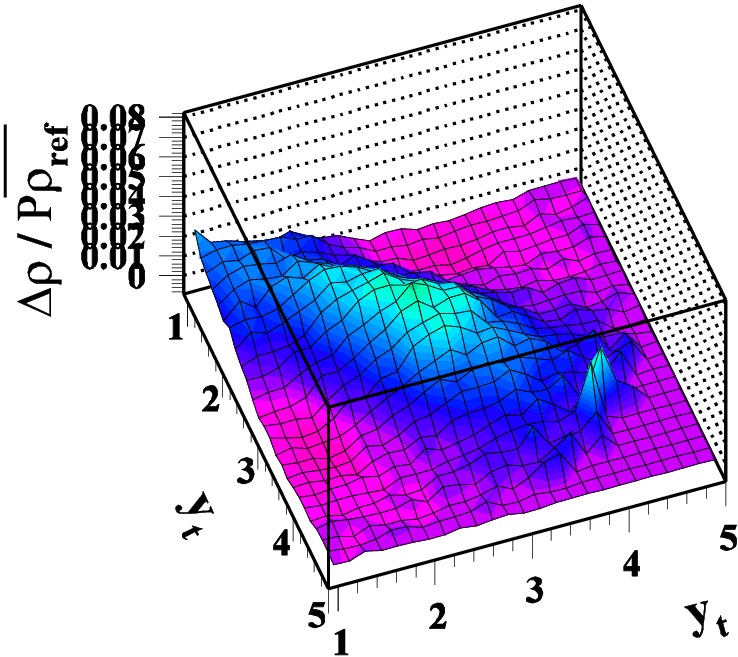}}
        \subfigure[AS, LS]{\includegraphics [width=0.245\textwidth]{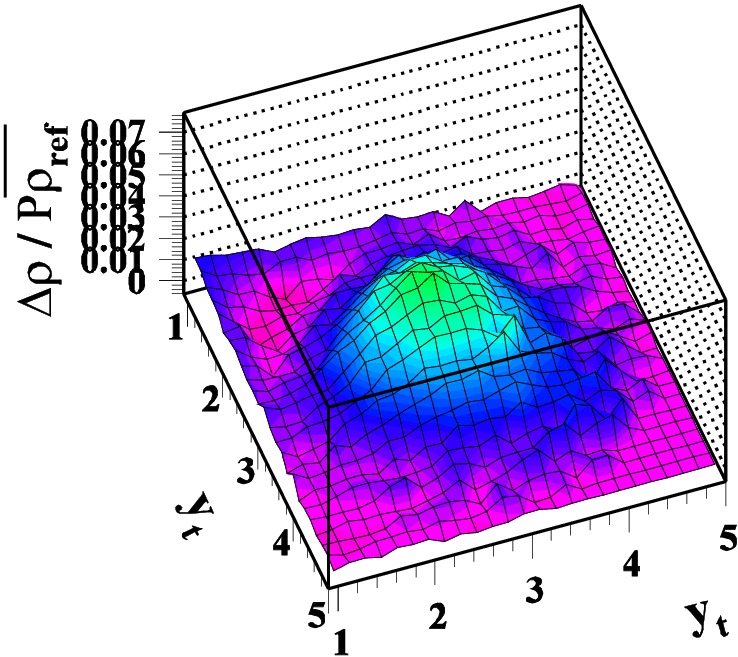}}
        \subfigure[AS, US]{\includegraphics [width=0.245\textwidth]{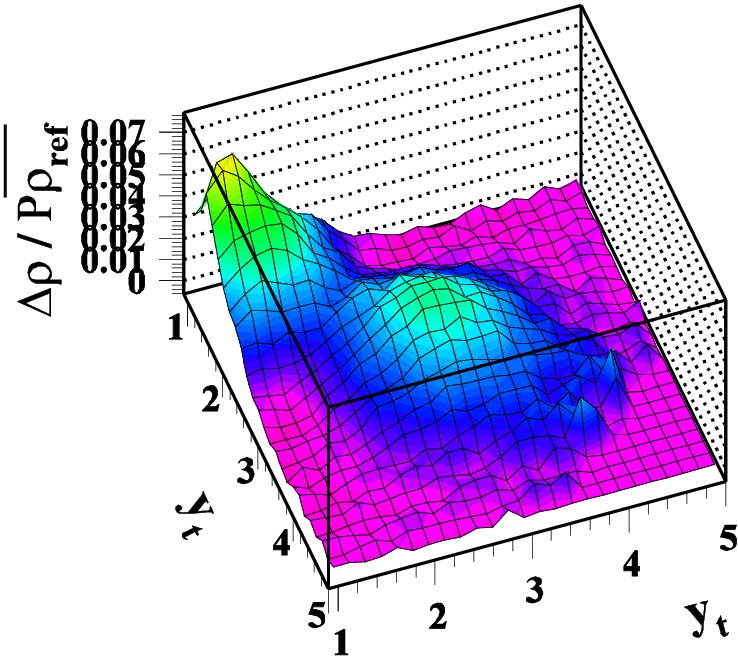}}
    \caption{Projections of two-particle correlations onto $(y_{t1},y_{t2})$ with cuts on angular space.
             The AS parton fragmentation region is independent of charge sign,
             the US pairs have additional soft component.
             The SS parton fragmentation region is restricted to US pairs.
    }
    \label{fig3}
\end{figure}

We examine $(y_{t1},y_{t2})$ in more detail by cutting on SS and AS pairs as
well as looking at LS and US pairs.
These four combinations are shown in Fig.~\ref{fig3}.
We refer to the region around $y_t \approx 3$ ($p_t \approx 1.5\rm\ GeV/c$) as the fragmentation region.
We see that for the AS this region is independent of the charge combination.
For SS pairs the fragmentation region is restricted to US pairs.
These correlations are dominated by low-$Q^2$ $\approx 3\rm\ GeV$ partons which primarily fragment into
hadron pairs, the pair being charge neutral.

\begin{figure}
    \subfigure[parton-fragment]{\includegraphics [width=0.19\textwidth]{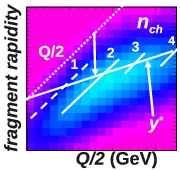}\label{fig3_1_a}}
    \subfigure[fragment-fragment]{\includegraphics [width=0.19\textwidth]{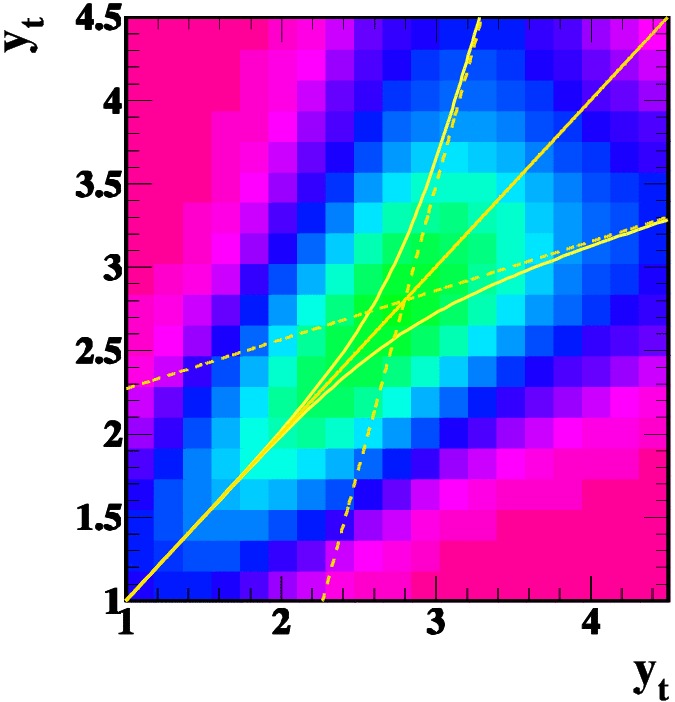}\label{fig3_1_b}}
    \subfigure[conditional cuts]{\includegraphics [width=0.19\textwidth]{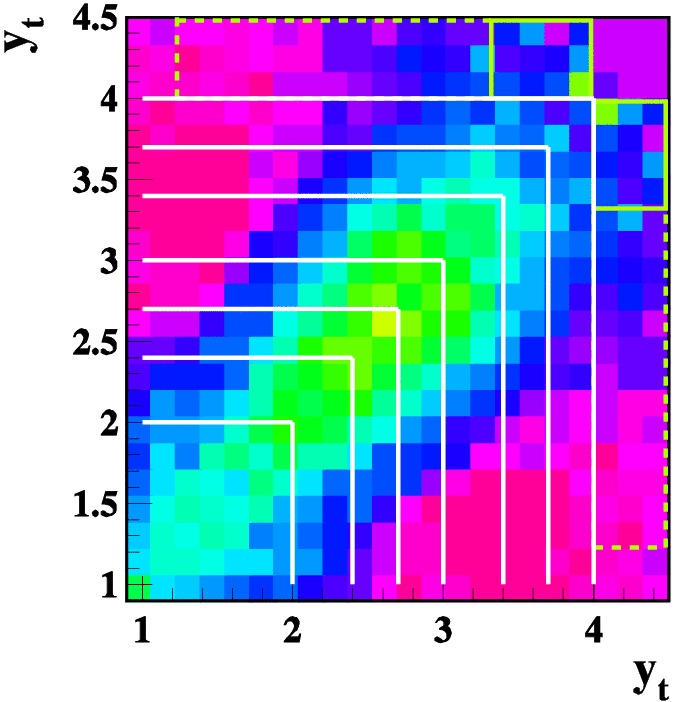}\label{fig3_1_c}}
    \subfigure[conditional slices and locus of modes]{\includegraphics [width=0.40\textwidth]{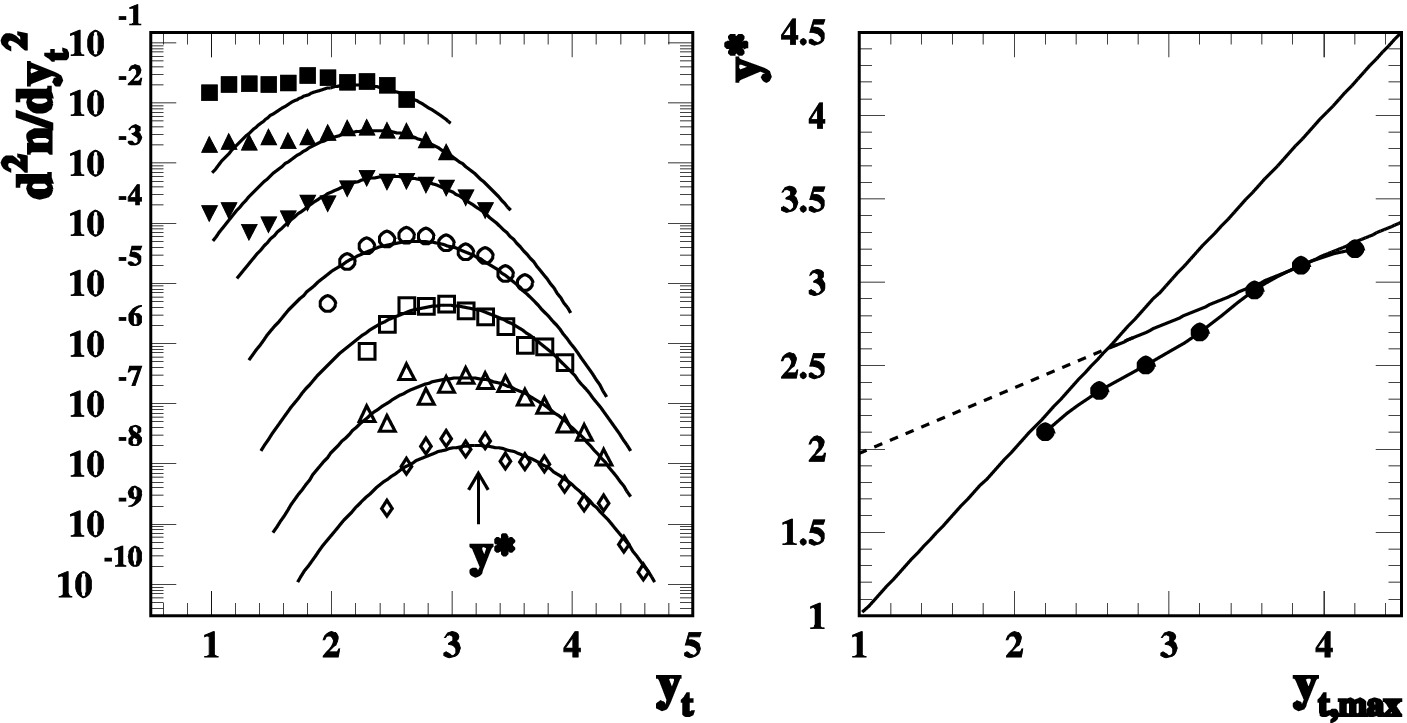}\label{fig3_1_de}}
    \caption{Panel~\protect\ref{fig3_1_a} shows the parton momentum versus fragment momentum.
             A vertical slice projected onto fragment momentum is a fragmentation function.
             In panel~\protect\ref{fig3_1_b} we symmetrize, integrating over parton momenta to get a figure
             we can compare to measurement.
             We note that at low momentum the fragment yield is dominated by partons fragmenting to two
             hadrons and is peaked along the diagonal.
             For larger momentum we can have parton to three or more fragments and we see the mode
             following the pQCD expectation.
             Panel~\protect\ref{fig3_1_c} is the measured SS $(y_{t1},y_{t2})$ correlation which should
             be compared with \protect\ref{fig3_1_b}.
             In panel~\protect\ref{fig3_1_de} we show conditional slices for the cuts defined in
             panel~\protect\ref{fig3_1_c}, fitting them to get the modes of the fragmentation functions.
             These closely follow the expectation sketched in panel~\protect\ref{fig3_1_b}.
    }
    \label{fig3_1}
\end{figure}

\begin{figure}
    \subfigure[``soft'' LS]{\includegraphics [width=0.245\textwidth]{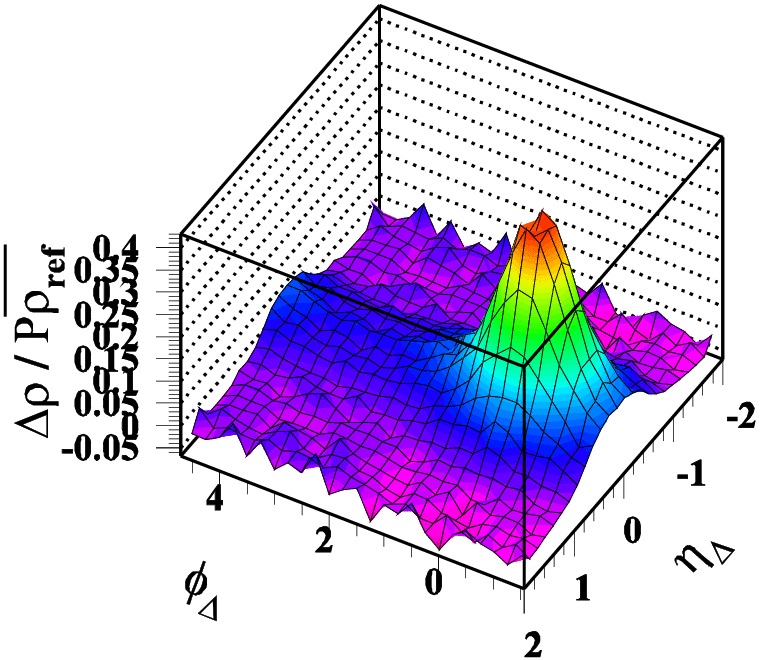}\label{fig4_a}}
    \subfigure[``soft'' US]{\includegraphics [width=0.245\textwidth]{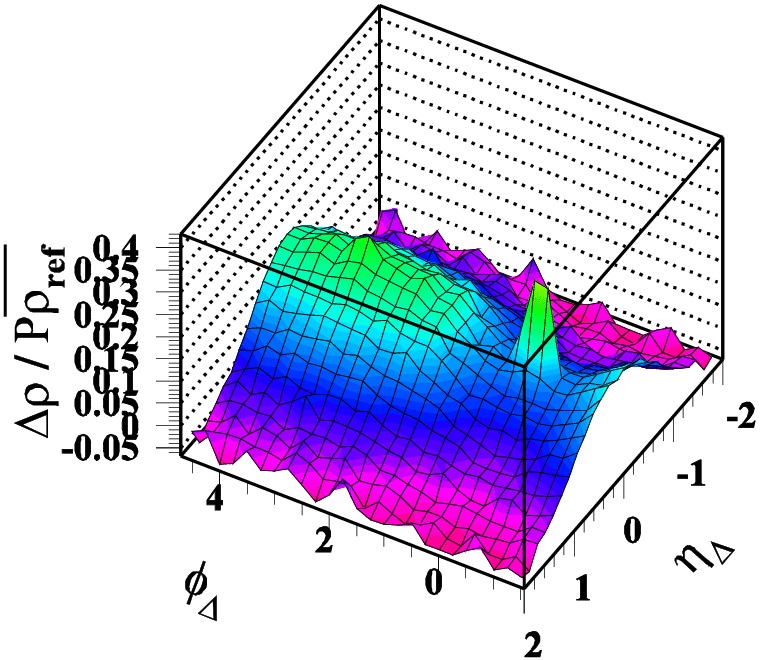}\label{fig4_b}}
    \subfigure[``hard'' LS]{\includegraphics [width=0.245\textwidth]{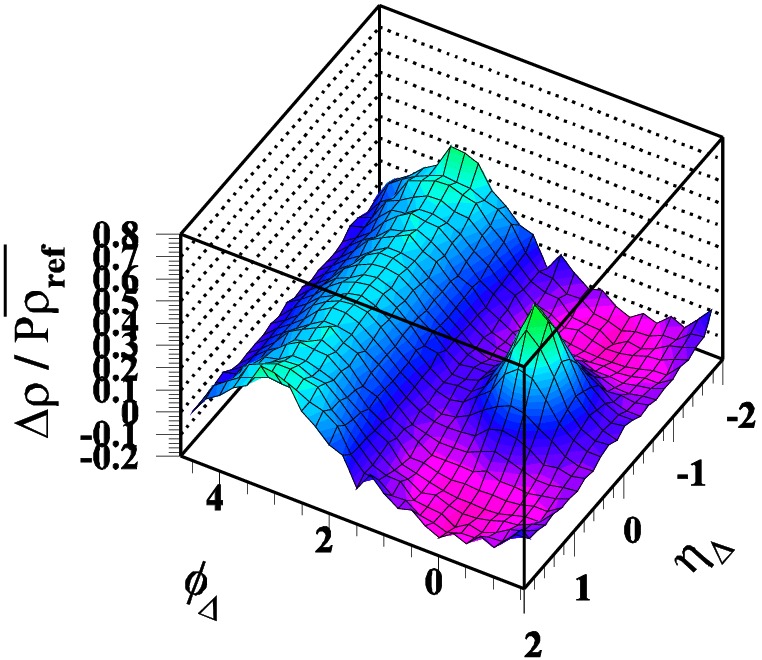}\label{fig4_c}}
    \subfigure[``hard'' US]{\includegraphics [width=0.245\textwidth]{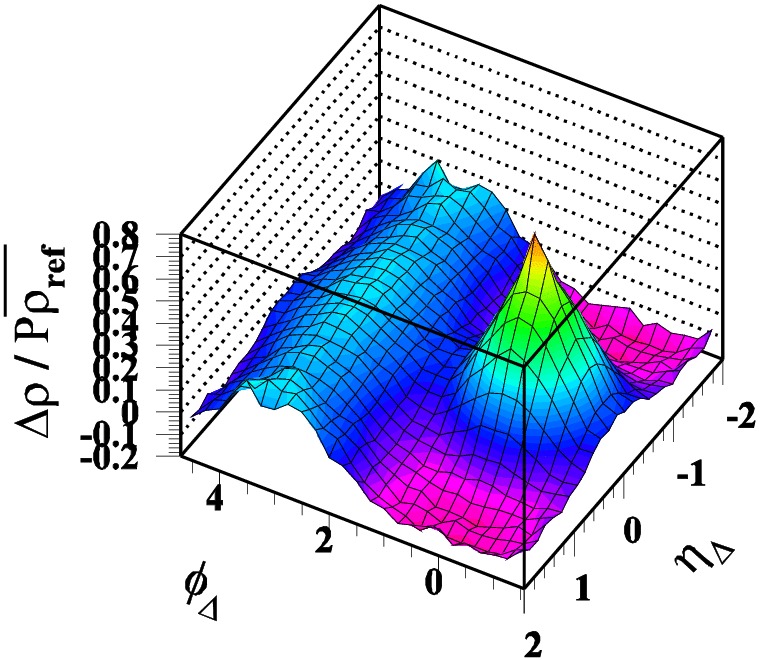}\label{fig4_d}}
    \caption{Projections of two-particle correlations onto $(\eta_\Delta,\phi_\Delta)$ space with cuts
             on $(y_{t1},y_{t2})$ space.
             Soft pairs have $y_{t1}+y_{t2} < 2.7$ and hard pairs $y_{t1},y_{t2} > 2$.
             For the soft pairs the LS has a strong HBT component while the
             US is dominated by local measure conservation such as momentum and charge.
             For the hard pairs the AS ridge is charge independent while the SS jet peak
             is primarily US.
    }
    \label{fig4}
\end{figure}

We have already seen in Fig.~\ref{fig2_a} that when projecting from the fragmentation region of $(y_{t1},y_{t2})$
the angular correlations have clear hard scattering (jet) structure.
We can also ask how the shape of the  $(y_{t1},y_{t2})$ correlations should look if it is really
due to fragmentation.
One can parameterize parton fragmentation functions (FFs) in a universal form~\cite{fragmentPaper}.
This allows us to construct a 2D plot with the parton momentum along one axis and the fragment momenta along the
other axis as shown in Fig.~\ref{fig3_1_a}.
A slice parallel to the fragment momentum axis is a fragmentation function.
The amplitude of these slices is determined by the underlying parton spectrum.
For large parton momentum the mode of the fragmentation function is well determined by pQCD~\cite{albino}\cite{fragmentPaper}.
As the parton energy decreases the number of hadrons it fragments to decreases until for some energy ($\approx 3\rm GeV$)
there are only two fragments (a parton may turn into a single hadron ``fragment'', but we are studying
two-particle correlations).
We do not observe the parton, only the hadrons.
So we symmetrize the parton versus fragment plot
to get a fragment versus fragment plot, integrating over parton momentum as shown in Fig.~\ref{fig3_1_b}.
At low $y_{t\Sigma}$ the yield is dominated by partons fragmenting to two hadrons and is peaked along
the diagonal.
As the fragment $y_t$ increases the importance of parton fragmentation to three, four and more particles increases.
The maximum in the $(y_{t1},y_{t2})$ plot deviates from the diagonal and follows the line of modes
predicted by pQCD.
This expectation is sketched in Fig.~\ref{fig3_1_b}.
Compare this to the measured $(y_{t1},y_{t2})$ correlation for SS fragments in Fig.~\ref{fig3_1_c}.

We can also go in the other direction, from the data to the line of modes.
We take the $(y_{t1},y_{t2})$ plot and project the conditional slices shown in Fig.~\ref{fig3_1_c} onto $y_t$.
For relatively large $y_t$ these are well described by a Gaussian with the same width
determined from the spectrum analysis~\cite{ppTwoComponent}.
Using this width to fit all $y_t$ slices works well down to $y_t$ of about 2.5 ($p_t \approx 1\rm\ GeV/c$).
A plot of the Gaussian centroid versus $y_t$ slice starts off along the diagonal
then curves, following the expected pQCD line of modes.
This is shown in Fig.~\ref{fig3_1_de} which should be compared with Fig.~\ref{fig3_1_b}.
The interpretation that the area around $y_t \approx 3$  is due to fragmentation
is supported not only by the shape of the angular correlations but also by details of measured
fragmentation functions.

We now turn to the angular correlation dependence on the $(y_{t1},y_{t2})$ space.
In Fig.~\ref{fig4} we show the angular correlations for LS and US pairs
for ``soft'' ($y_{t\Sigma} < 2.7$) and ``hard'' ($y_{t1},y_{t2} > 2$) pairs.
We see a pronounced HBT signal in the LS pairs at (0,0), especially for the soft cut.
The largest component of the soft US pairs is a 1D Gaussian on $\eta_\Delta$
due to charge ordering in projectile nucleon fragmentation along the beam axis.
Imposed on this is a dip centered at $(0,0)$ due to momentum conservation (when there are
only a few particles in an event they will be biased against having the same direction, unless
they are fragments from a single hard-scattered parton) and a narrow peak at $(0,0)$ 
due to $e^+e^-$ from $\gamma$-induced pair production.
The hard pairs have an AS ridge that is independent of charge,
the SS peak is primarily in US pairs.

\begin{figure}
    \subfigure[$\sigma_\eta$ and $\sigma_\phi$ widths of SS peak.]{\includegraphics [width=0.63\textwidth]{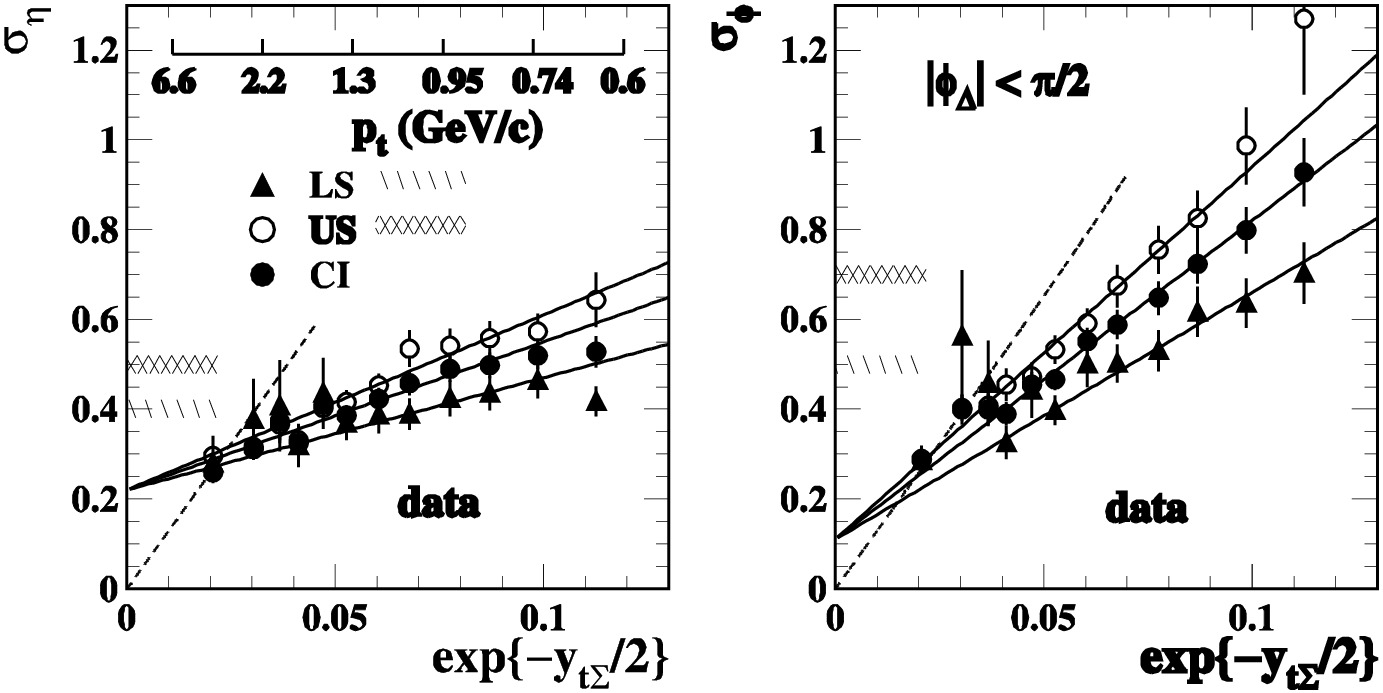}\label{fig5_a}}
    \subfigure[SS peak for $y_{t\Sigma} \approx 3.5$]{\includegraphics [width=0.3\textwidth]{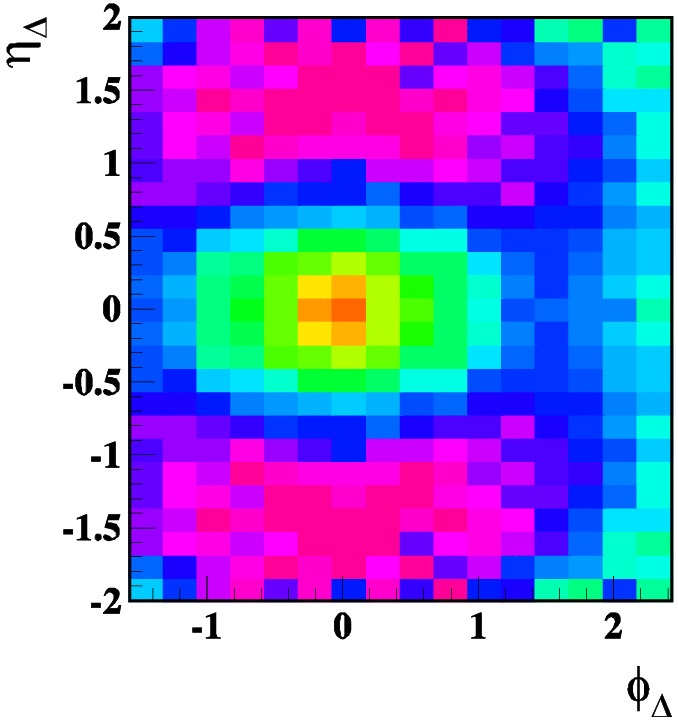}\label{fig5_b}} \\
    \subfigure[$j_{t\eta}$ and $j_{t\phi}$]{\includegraphics [width=0.64\textwidth]{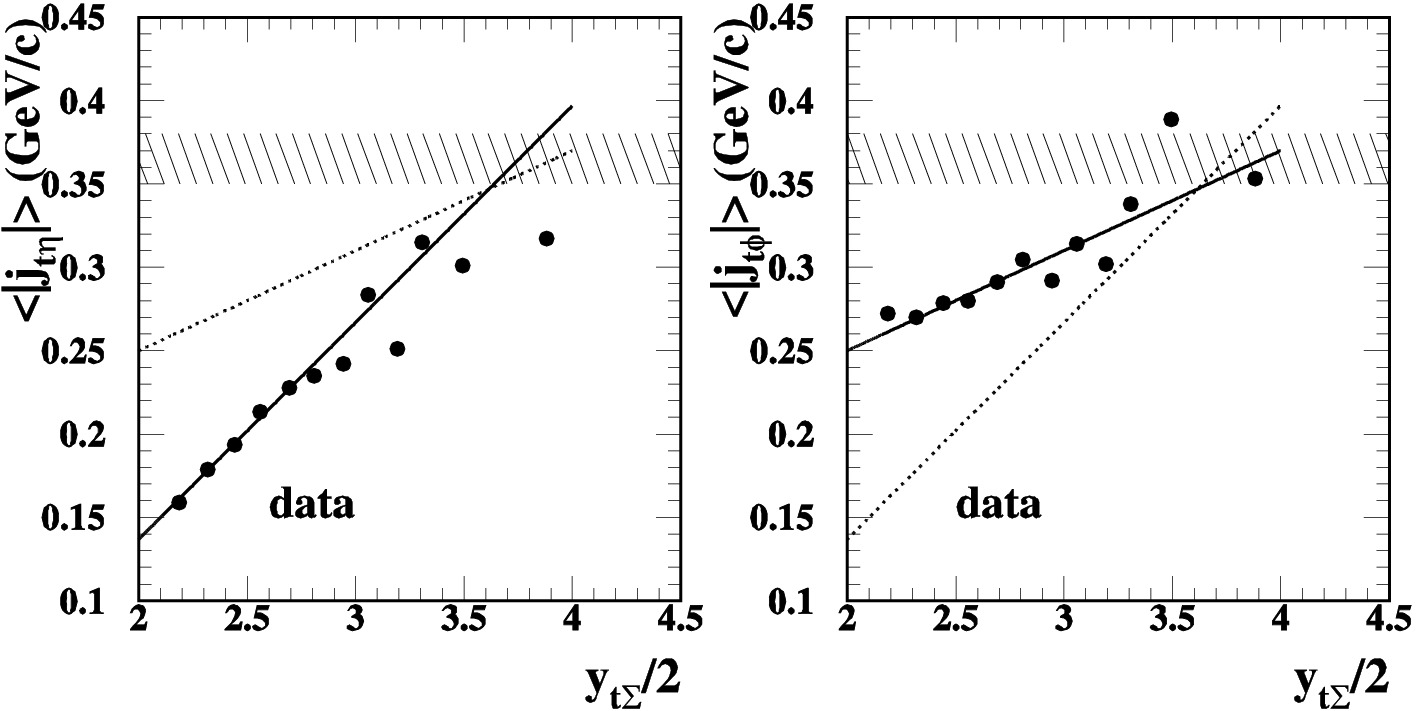}\label{fig5_c}}
    \subfigure[SS peak for $y_{t\Sigma} \approx 6.5$]{\includegraphics [width=0.3\textwidth]{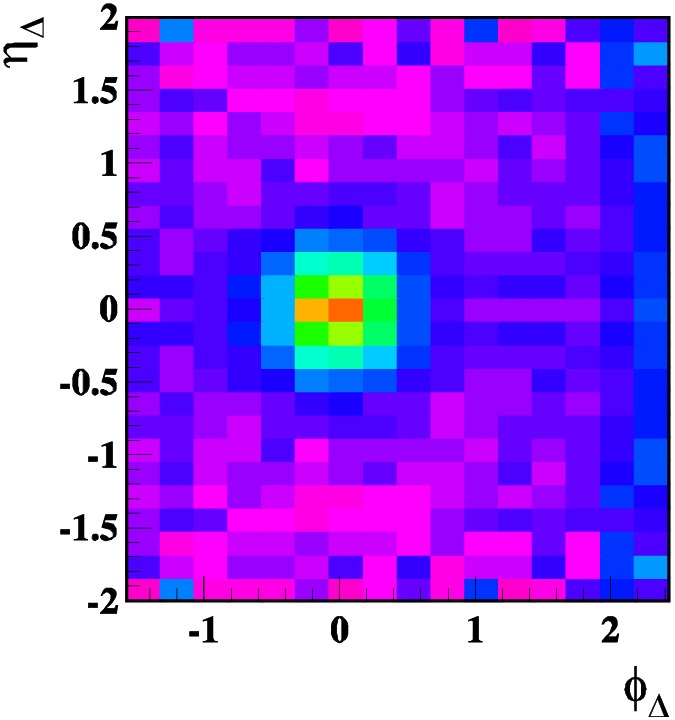}\label{fig5_d}}
    \caption{Panel~\protect\ref{fig5_a} shows the fitted widths of the SS peak, the x-axis being
             essentially $1/p_t$ of the pair.
             The widths depend on the direction ($\eta$ or $\phi$) and the charge combination but are
             approximately linear in this variable.
             Panel~\protect\ref{fig5_c} shows the extracted $j_t$ values for the $\eta$ and $\phi$ directions,
             having very different values for low $y_t$ but approaching the $j_t$ scaling value at high $y_t$.
             Panels~\protect\ref{fig5_b} and \protect\ref{fig5_d} are projections onto $(\eta_\Delta,\phi_\Delta)$ for
             low and high $y_{t\Sigma}$ cuts and plotted in 1:1 aspect ratio and graphically showing
             the change in the shape of the SS peak as a function of pair momentum.
    }
    \label{fig5}
\end{figure}

In Fig.~\ref{fig5} we examine this jet structure in more detail by making finer cuts on $y_{t\Sigma}$.
As $y_{t\Sigma}$ is increased the SS becomes sharper and the AS ridge becomes narrower.
The width of the AS ridge is affected by $k_t$, a measure of the initial-state
parton transverse momenta.
The widths of the SS 2D Gaussian are related to
$j_{t\eta}$ and $j_{t\phi}$, measures of the transverse momentum with respect to the thrust axis of
the fragmentation process~\cite{feynmanKt}\cite{jtScaling}.
The relations between the widths and the $j_t$s and $k_t$ is usually derived for the asymmetric
case of a leading particle~\cite{jJiaKin}\cite{jRakKin}.
In this analysis we use a formulation symmetric in the two particles~\cite{tTrainorKin}.
In addition, since the most-probable parton has $Q\approx 3\rm\ GeV$ and the observed $k_t$ is typically
around 1~GeV/c (with the $j_t$s being not too much smaller) we cannot use small-angle approximations.
When $y_{t\Sigma}$ is comparable to $j_t$ or $k_t$ the fragment distributions will be affected by kinematic limits.
The results of fits to the SS peak are shown in Fig.~\ref{fig5}.
Panel~\ref{fig5_a} shows the widths in the $\eta_\Delta$ and $\phi_\Delta$ directions.
Since $y_t$ is the log of $p_t$ the x-axis is essentially $1/p_t$,
the $p_t$ of the pair decreasing from left to right.
We see that for this variable the $\eta$ and $\phi$ widths of the SS peak are approximately linear,
but different from each other and dependent on charge combination.
Panel~\ref{fig5_c} shows the inferred $j_t$s, this time plotted on $y_{t\Sigma}$.
The difference in the $j_{t\eta}$ and $j_{t\phi}$ is greatest at low $y_{t\Sigma}$.
As $y_{t\Sigma}$ is increased they both increase until at large enough momentum they approach
the observed perturbative $j_t$-scaling value~\cite{jtScaling}.

\begin{figure}
    \subfigure[AS $(y_{t1},y_{t2})$]{\includegraphics [width=0.3\textwidth]{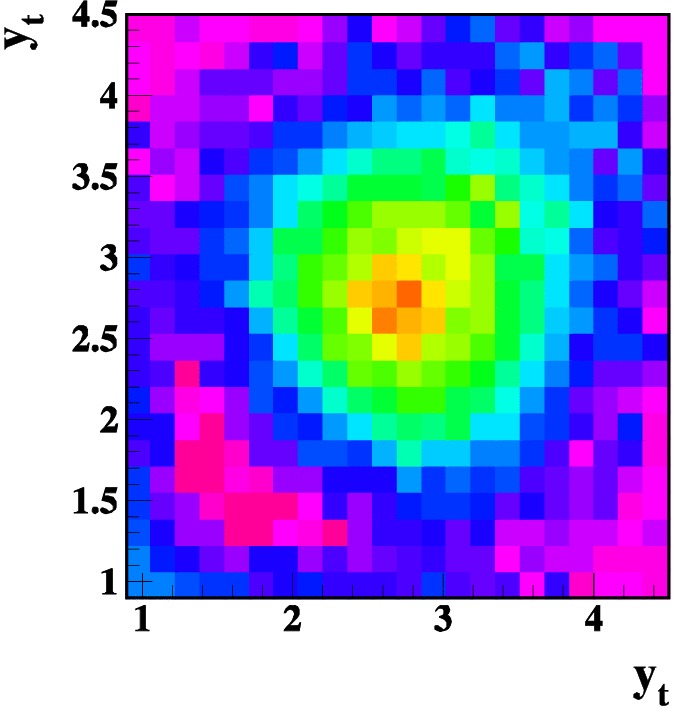}\label{fig6_a}}
    \subfigure[SS $(y_{t1},y_{t2})$]{\includegraphics [width=0.3\textwidth]{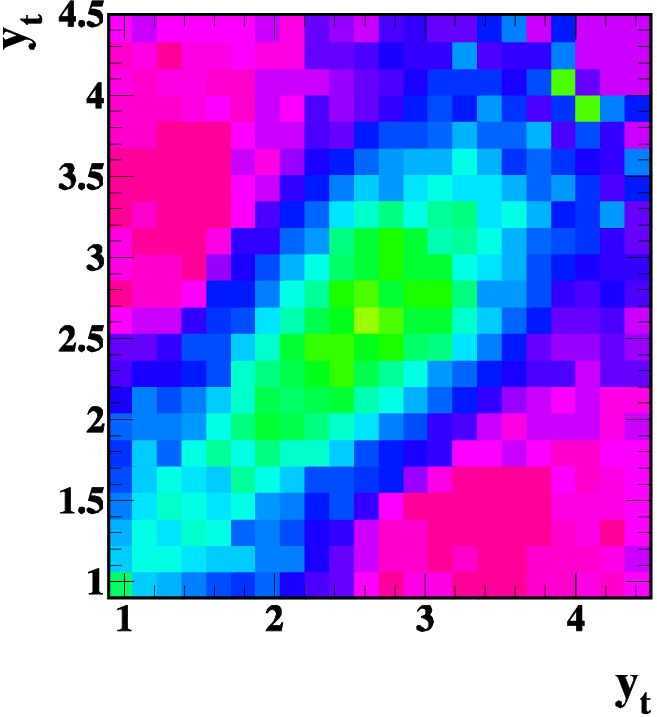}\label{fig6_b}}
    \subfigure[``hard'' $(\eta_\Delta,\phi_\Delta)$]{\includegraphics [width=0.3\textwidth]{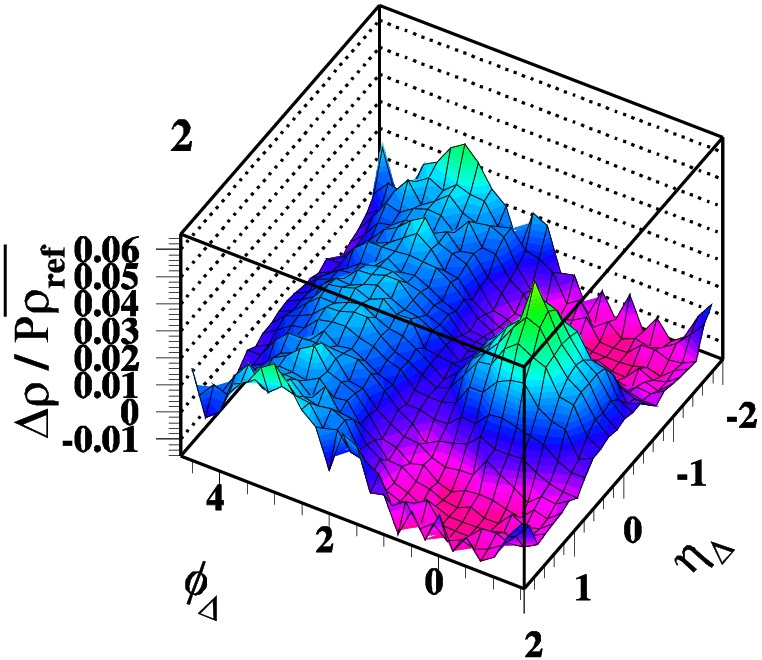}\label{fig6_c}}
    \caption{Effect of $k_t$ on $(y_{t1},y_{t2})$ correlations.
             Panel~\protect\ref{fig6_a} shows the AS $(y_{t1},y_{t2})$, broadened when the
             $k_t$ of the two scattered partons are parallel to each other and the direction
             of the momentum transfer.
             The SS correlation, shown in panel~\protect\ref{fig6_b}, is not affected by $k_t$.
             Panel~\protect\ref{fig6_c} indicates how one can extract $k_t$ by measuring the
             width of the AS ridge.
    }
    \label{fig6}
\end{figure}

The width of the AS ridge can be used to infer $k_t$, which has components from intrinsic
initial-state parton momenta and initial-state radiation~\cite{tannenbaum}.
The momentum transfer between the scattered partons defines a direction $\vec{q}$.
When the initial $k_t$ of both partons is perpendicular to $\vec{q}$ but parallel to each other the
AS ridge width is widest.
The effects of $k_t$ also show up in the AS $(y_{t1},y_{t2})$ correlations.
When the $k_t$s of the two partons are parallel to each other and parallel to $\vec{q}$
the $y_t$ values for the fragments of one of the partons will be boosted while the
$y_t$ for fragments of the other parton are reduced.
This populates the off-diagonal region of the $(y_{t1},y_{t2})$ correlation.
We show the AS $(y_{t1},y_{t2})$ correlation in panel~\ref{fig6_a}.
In contrast, the SS correlation, shown in panel~\ref{fig6_b},  is unaffected by $k_t$
and is considerably narrower.
Typically, measured $k_t$ is about 1~GeV/c, nearly the same momentum as the minijet partons
we are probing.
It may be possible in some cases for the $k_t$ to boost the scattered partons enough so that
they both actually emerge on the same side.

We have seen how we can use standard correlation measures to extract information
in p-p collisions.
We find clear signatures of hard parton scattering at surprisingly low parton momenta.
We have also seen that there is a nice correspondence between two-particle correlations
and a spectrum analysis which indicates a hard component appearing in perhaps 1\% of
NSD p-p collisions at 200~GeV~\cite{ppTwoComponent}.
We want to use this information as the baseline when analyzing A-A collisions, keeping in mind
that the fraction of particles produced by hard scattering depends on centrality.
So the superposition of N-N collisions should be weighted appropriately to account for the
hard p-p spectrum component.

\section{Two-particle correlations in A-A collisions}

\begin{figure}
    \subfigure[84-93\%]{\includegraphics [width=0.245\textwidth]{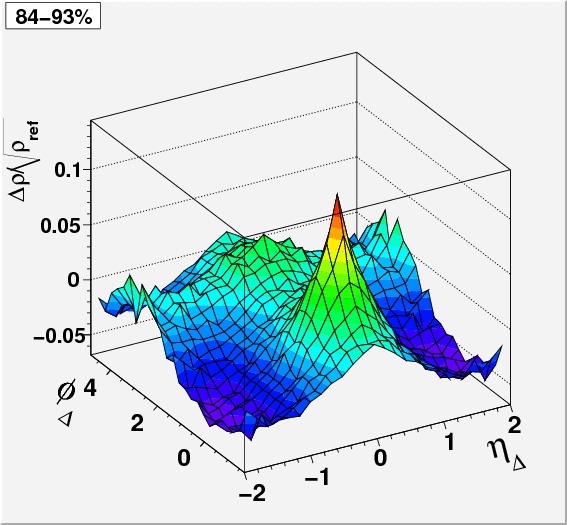}\label{fig7_a}}
    \subfigure[55-64\%]{\includegraphics [width=0.245\textwidth]{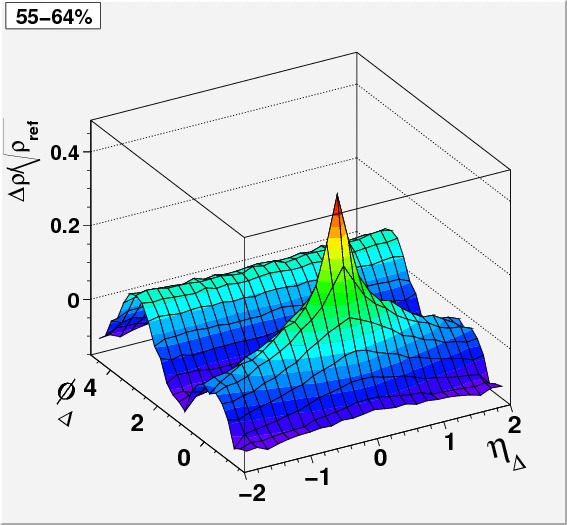}\label{fig7_b}}
    \subfigure[46-55\%]{\includegraphics [width=0.245\textwidth]{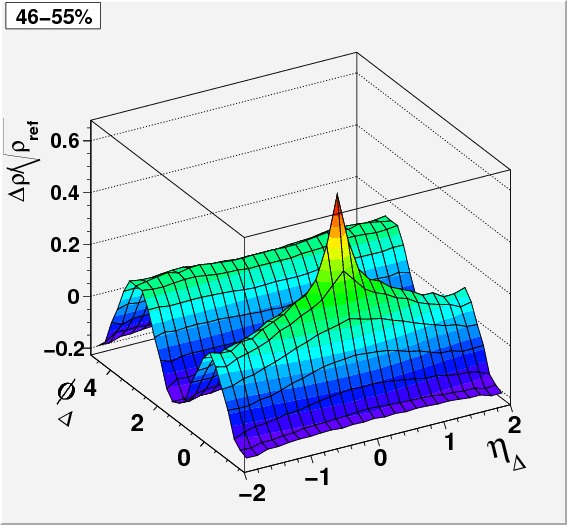}\label{fig7_c}}
    \subfigure[5-9\%]{\includegraphics [width=0.245\textwidth]{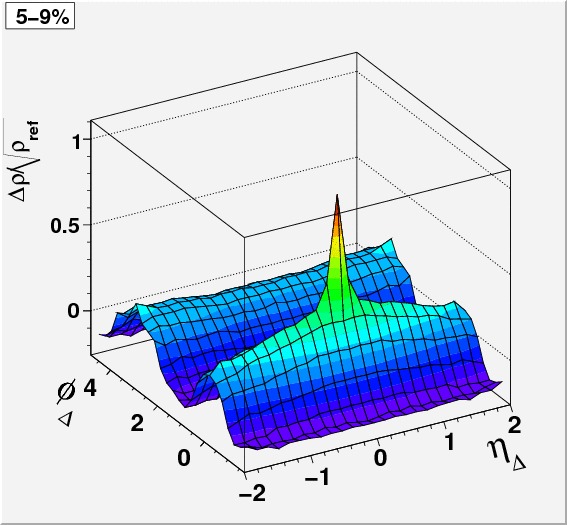}\label{fig7_d}}
    \caption{Samples of $(\eta_\Delta,\phi_\Delta)$ correlations for 200~GeV Au-Au collisions.
             Panel~\protect\ref{fig7_a} is most peripheral (84-94\%) while panel~\protect\ref{fig7_b} is nearly central.
             The middle two panels are on either side of the sharp transition, 55-64\% and 46-55\% centralities.
    }
    \label{fig7}
\end{figure}

In this paper we only discuss angular correlations from symmetric A-A collisions.
We have analyzed Au-Au collisions at 200~GeV and 62~GeV as well as Cu-Cu
collisions at the same energies.
We see evolution with centrality from a correlation structure consistent with
p-p collisions for the more peripheral events to a structure dominated by the
SS 2D peak being strongly elongated along $\eta_\Delta$ and an AS ridge consistent with
back-to-back hard parton scattering.
For mid-central collisions there is in addition a $\cos(2\phi_\Delta)$ quadrupole component.
Examples for a few centralities of 200~GeV Au-Au collisions are shown in Fig.~\ref{fig7}.
The other collision systems are qualitatively similar.

We quantitatively describe the 2D correlation structure with a SS 2D Gaussian (with different $\eta_\Delta$
and $\phi_\Delta$ widths), a $\cos(\phi_\Delta-\pi)$ dipole describing the AS ridge and a 1D Gaussian
on $\eta_\Delta$~\cite{zyamTrainor}. 
To complete the description of p-p collisions we include a narrow exponential at the origin
to describe $e^+e^-$ pairs from $\gamma$ conversions as well as an overall offset.
We don't infer any physics from the exponential peak or the offset.
This model function with ten parameters works well for the most peripheral and fairly
well for the most central data.
For intermediate centralities a $\cos(2\phi_\Delta)$ term is required as well.

\begin{figure}
  \begin{center}
    \includegraphics [width=0.8\textwidth]{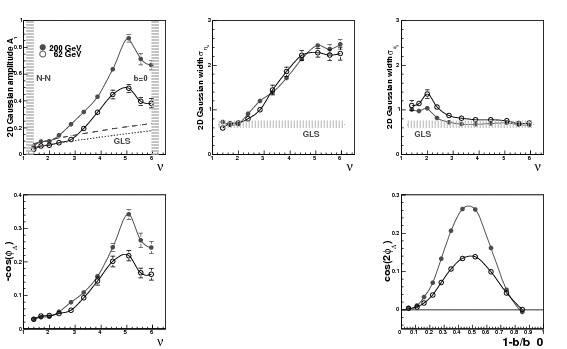}
  \end{center}
    \caption{Values of fit parameters for Au-Au collision systems.
             For the more peripheral data the SS peak amplitude follows
             the binary scaling expectation, exceeding it dramatically at a transition centrality
             that depends on the energy/collision system.
             The AS ridge follows this trend very closely and the $\eta_\Delta$ width
             of the SS peak also increases substantially at a similar centrality.
             In contrast the $\phi_\Delta$ width of the SS peak decreases with increasing
             centrality and the $\cos(2\phi_\Delta)$ term has an interesting but smooth evolution,
             already having a significant value before the transition centrality.
    }
    \label{fig8}
\end{figure}

Fit parameters are shown in Fig.~\ref{fig8}.
The most striking feature of the fit parameters is that the SS Gaussian amplitude
closely follows the expectation for binary scaling up to some centrality then greatly exceeds it.
The AS ridge described by $\cos(\phi_\Delta-\pi)$ tracks the amplitude of the SS Gaussian.
The $\eta_\Delta$ width of the SS peak also increases greatly close to the same centrality.
In contrast to the $\eta_\Delta$ width, the $\phi_\Delta$ width starts at the same value as
we find for p-p collisions but then actually decreases.
In contrast to the sharp transition of the jet amplitude the $\cos(2\phi_\Delta)$ component
has an {\em interesting but unrelated} evolution on centrality.

\begin{figure}
    \subfigure[84-93\%]{\includegraphics [width=0.245\textwidth]{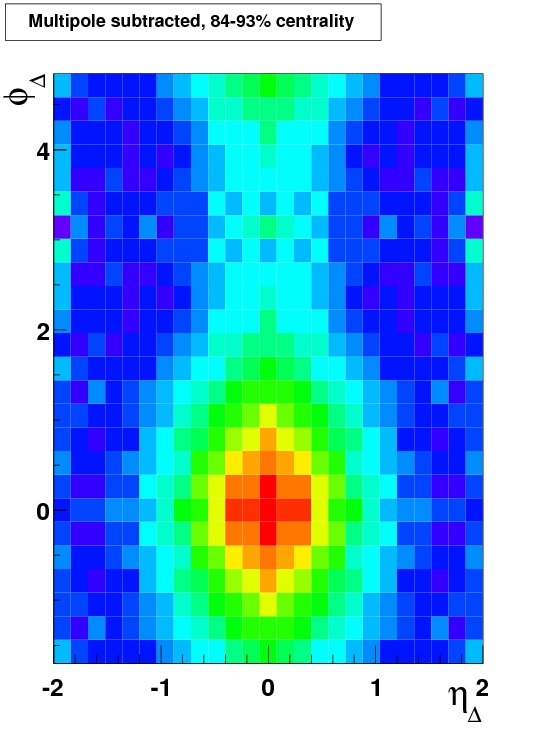}\label{fig9_a}}
    \subfigure[55-64\%]{\includegraphics [width=0.245\textwidth]{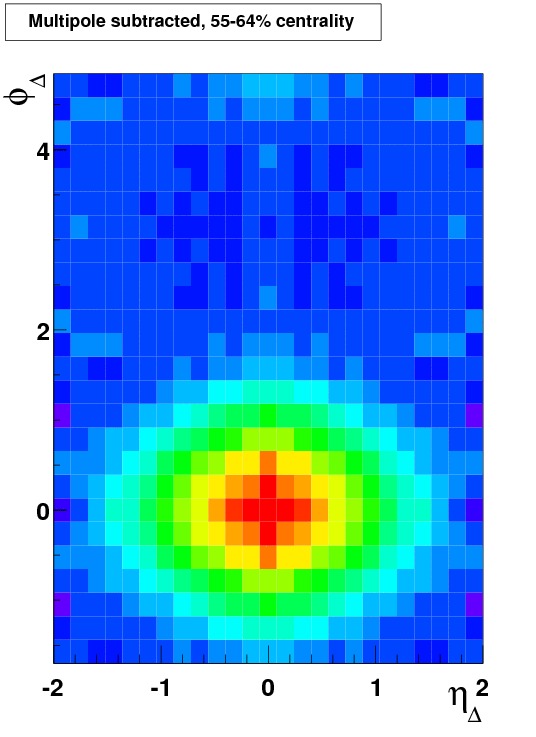}\label{fig9_b}}
    \subfigure[46-55\%]{\includegraphics [width=0.245\textwidth]{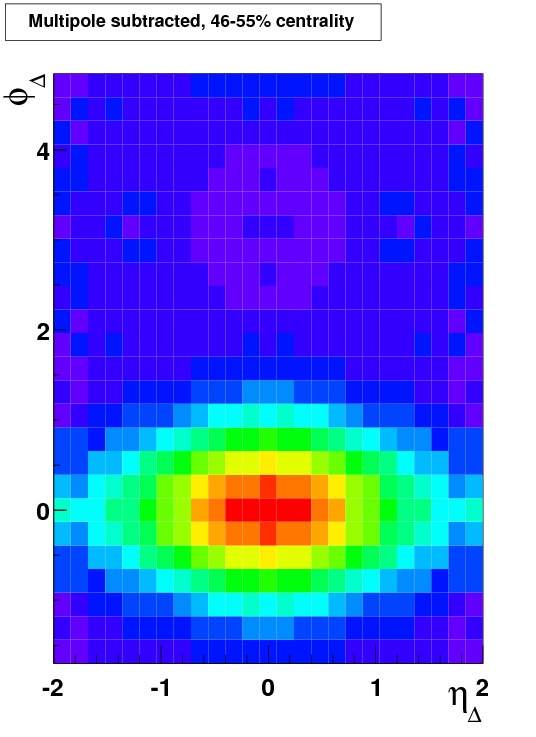}\label{fig9_c}}
    \subfigure[5-9\%]{\includegraphics [width=0.245\textwidth]{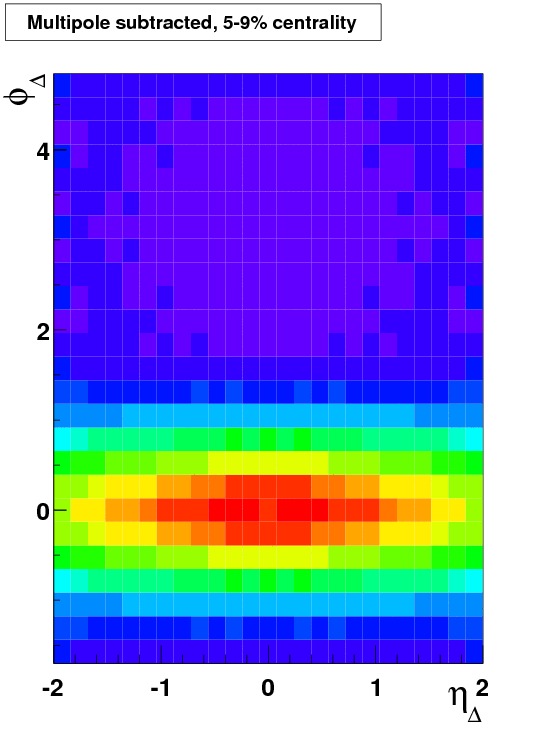}\label{fig9_d}}
    \caption{Samples of ($\eta_\Delta,\phi_\Delta)$ correlations for 200~GeV Au-Au collisions
             after multipoles and sharp peak have been subtracted and plotted in a 1:1 aspect ratio.
             Panel~\protect\ref{fig9_a} is most peripheral (84-94\%) while panel~\protect\ref{fig9_b} is nearly central.
             The middle two panels are on either side of the sharp transition, 55-64\% and 46-55\% centralities.
    }
    \label{fig9}
\end{figure}

We emphasize the deformation of the SS Gaussian peak by plotting the data after subtracting
the multipoles determined by the fit (we also subtract the sharp exponential peak)
in Fig.~\ref{fig9}, plotting the data in a one-to-one aspect ratio.
The most peripheral bin is actually elongated along the $\phi_\Delta$ direction.
But as we increase centrality the peak becomes symmetric, then within a narrow centrality range becomes
greatly elongated along $\eta_\Delta$.

\begin{figure}
    \subfigure[84-93\%]{\includegraphics [width=0.245\textwidth]{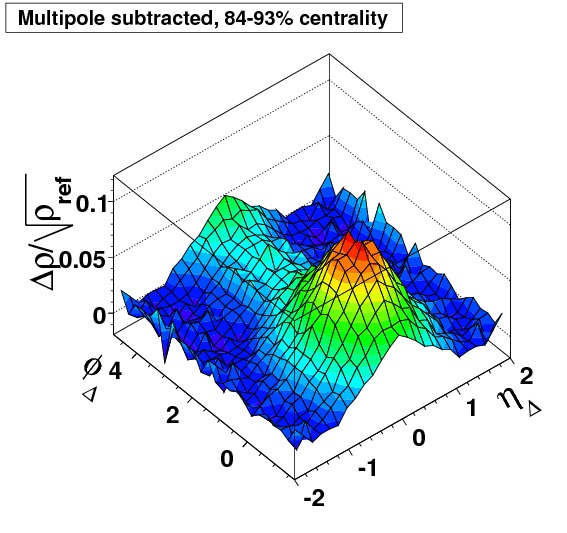}\label{fig9_1_a}}
    \subfigure[55-64\%]{\includegraphics [width=0.245\textwidth]{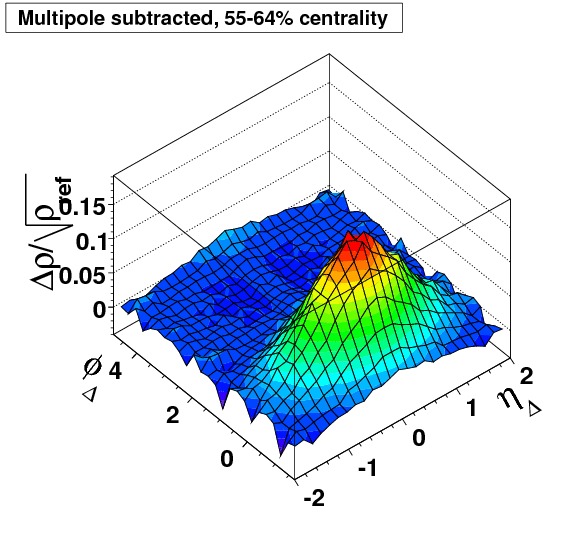}\label{fig9_1_b}}
    \subfigure[46-55\%]{\includegraphics [width=0.245\textwidth]{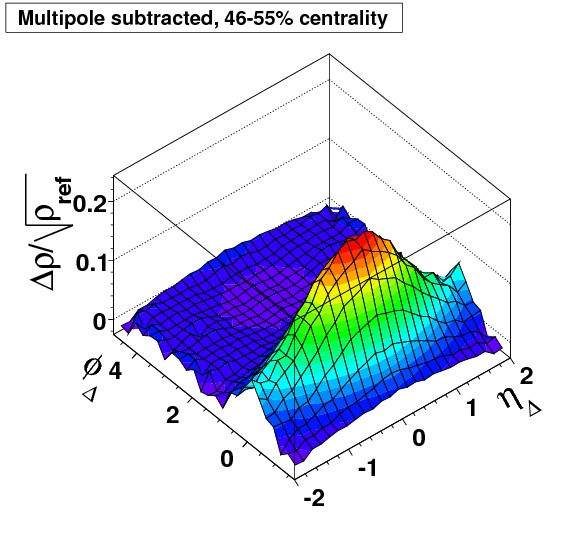}\label{fig9_1_c}}
    \subfigure[5-9\%]{\includegraphics [width=0.245\textwidth]{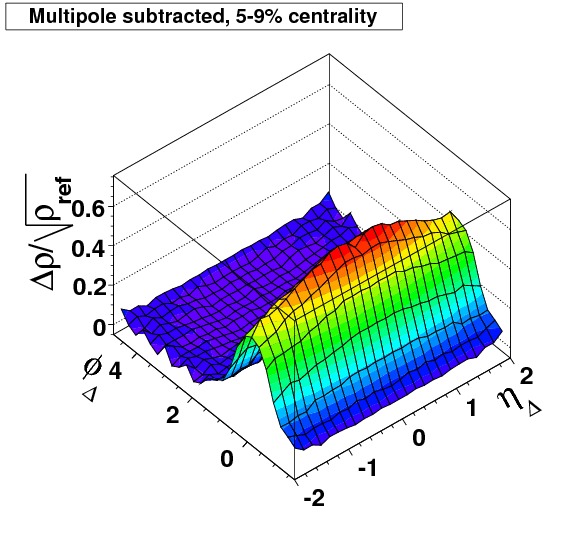}\label{fig9_1_d}}
    \caption{The same centralities as Fig.~\protect\ref{fig9} but plotted in a view similar to
             Fig.~\protect\ref{fig7}.
             We have subtracted the fitted dipole, quadrupole and sharp exponential peak from the data
             and rotated slightly.
             We see the dipole and quadrupole terms have exhausted the $\phi_\Delta$ dependent
             structure on the AS.
    }
    \label{fig9_1}
\end{figure}

In Fig.~\ref{fig9_1} we present an isometric view of the same data.
We subtract the fitted dipole, quadrupole and sharp exponential peak from the data.
The sharp peak affects only a few bins around the origin.
Our parameterization of the data does a good job describing the $\phi_\Delta$ structure of the AS peak.
There is a small but statistically significant $\eta_\Delta$ structure remaining on the AS.
We note that the $\phi_\Delta$ dependence of the SS 2D peak plotted here can be decomposed into
Fourier components on azimuth and would contribute significantly to an inferred quadrupole component.
We believe this peak is due to hard scattering (it certainly is for peripheral collisions)
and in any case this shape cannot be due to a medium response from a pressure gradient.

\section{$p_t$ correlations}

\begin{figure}
    \subfigure[84-93\%]{\includegraphics [width=0.245\textwidth]{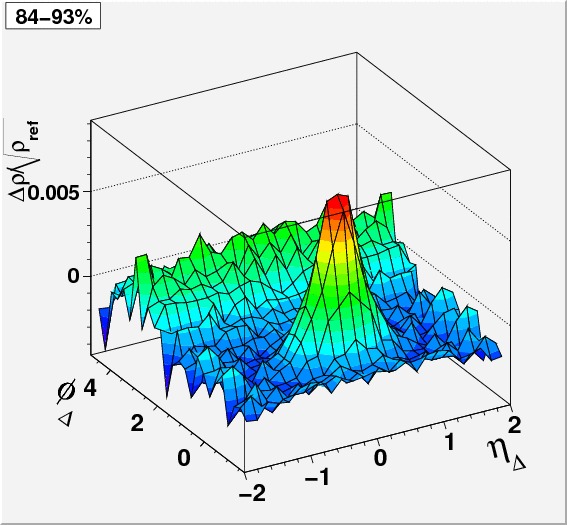}\label{fig10_a}}
    \subfigure[55-64\%]{\includegraphics [width=0.245\textwidth]{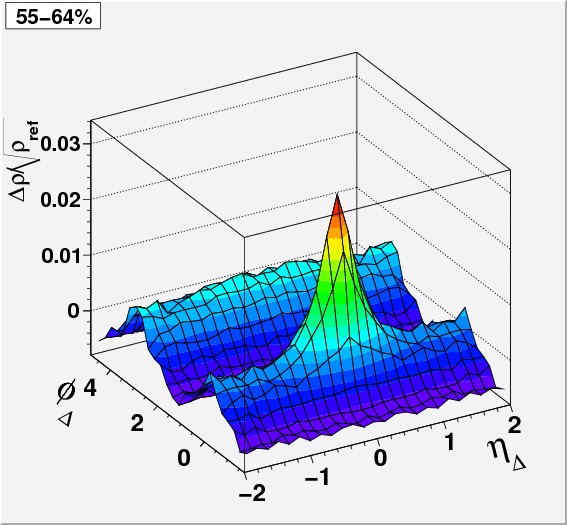}\label{fig10_b}}
    \subfigure[46-55\%]{\includegraphics [width=0.245\textwidth]{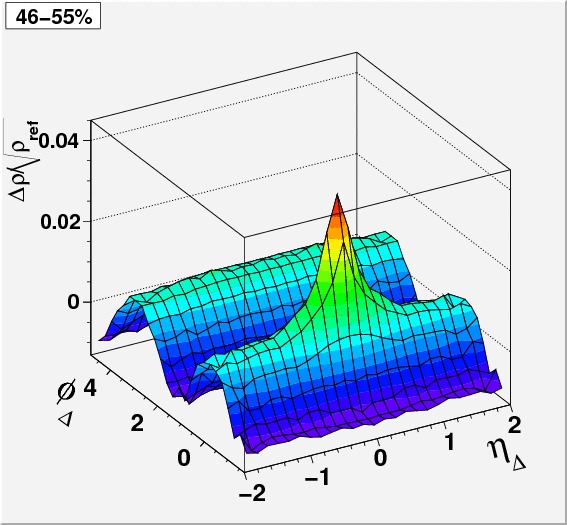}\label{fig10_c}}
    \subfigure[5-9\%]{\includegraphics [width=0.245\textwidth]{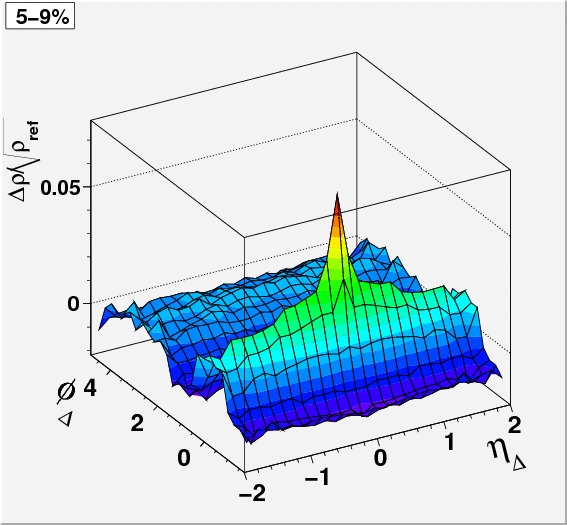}\label{fig10_d}}
    \caption{Samples of $(\eta_\Delta,\phi_\Delta)$ $p_t$ correlations for 200~GeV Au-Au collisions.
             These are the same centralities shown in Fig.~\protect\ref{fig7}.
    }
    \label{fig10}
\end{figure}

We now return to the topic of $p_t$ correlations which we mentioned briefly in section~1.
For number correlations the Pearson's correlation coefficient quantifies how the number density
at two locations vary with respect to each other.
One could also ask how the total momentum densities are correlated.
This turns out to be still dominated by number correlations.
We instead ask how the mean $p_t$s are correlated.
Specifically,
\begin{equation*}
    \frac{\Delta\rho}{\sqrt{\rho_{ref}}} = \frac{\overline{(p_t-n\hat{p_t})_a (p_t-n\hat{p_t})_b}}{\sqrt{\bar{n}_a\bar{n}_b}},
\end{equation*}
where $\hat{p_t}$ is the mean $p_t$ of the parent population.
We show examples of these correlations in Fig.~\ref{fig10}.
There is a qualitative similarity with the number correlations but there are quantitative differences.
Even the most peripheral bin shows no indication of a 1D Gaussian on $\eta_\Delta$.
The sharp peak at $(0,0)$ is dominated by HBT, opposed to number correlations where 
$e^+e^-$ is a significant contribution.
The broader peak centered at $(0,0)$ is a 2D Gaussian for number correlations but has a catenary
shape on $\eta_\Delta$ for $p_t$ correlations.

We previously looked at mean-$p_t$ correlations when we were measuring $p_t$ fluctuations and found
we could invert fluctuation scale dependence to obtain
detector-acceptance-independent correlations~\cite{fluctInvPaper}.
We found that the SS 2D peak amplitude was nearly proportional to the mean participant path length ($\nu$)
from the most peripheral to about 30\% most-central, falling for the most central.
As the centrality selection increased the SS 2D $\eta_\Delta$ width increased by about 60\%
while the $\phi_\Delta$ width decreased by about 30\%.
These trends are qualitatively similar to number correlations but quantitatively very different.
In addition, there appears to be a ``recoil hole'' around the SS peak~\cite{fluctInvPaper} which is
not observed in number correlations.
We also found that although Hijing does produce a SS 2D peak and an AS ridge
in $p_t$ correlations these are nearly centrality independent as well as having the
wrong detailed shape, with no indication of a recoil hole~\cite{fluctInvPaper}.
We are presently working on a detailed description of these correlations
to characterize our direct $p_t$ correlation measurements.

\begin{figure}
    \subfigure[80-90\%]{\includegraphics [width=0.245\textwidth]{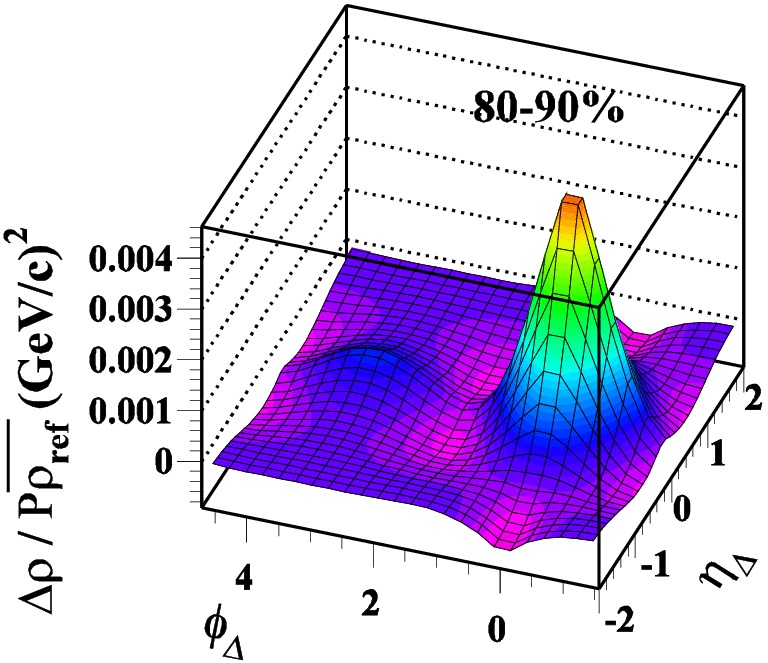}\label{fig11_a}}
    \subfigure[45-55\%]{\includegraphics [width=0.245\textwidth]{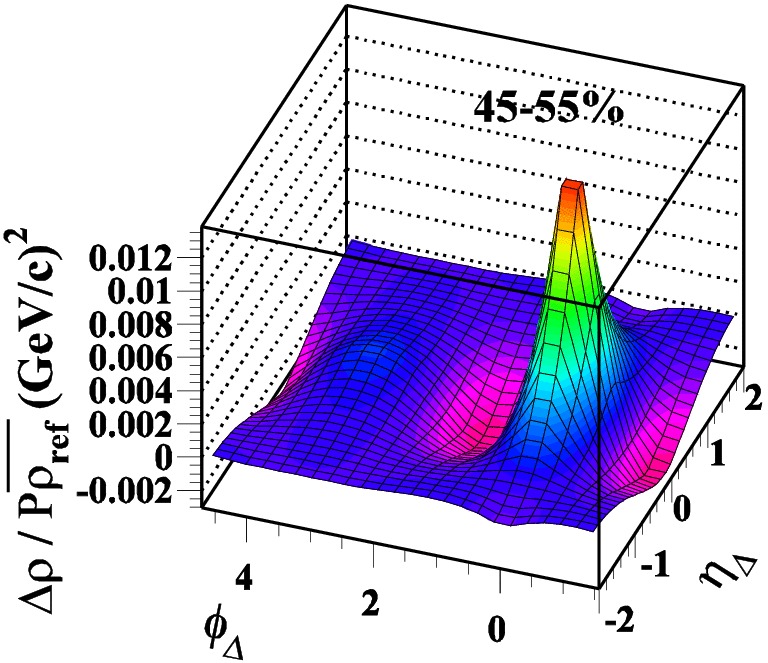}\label{fig11_b}}
    \subfigure[20-30\%]{\includegraphics [width=0.245\textwidth]{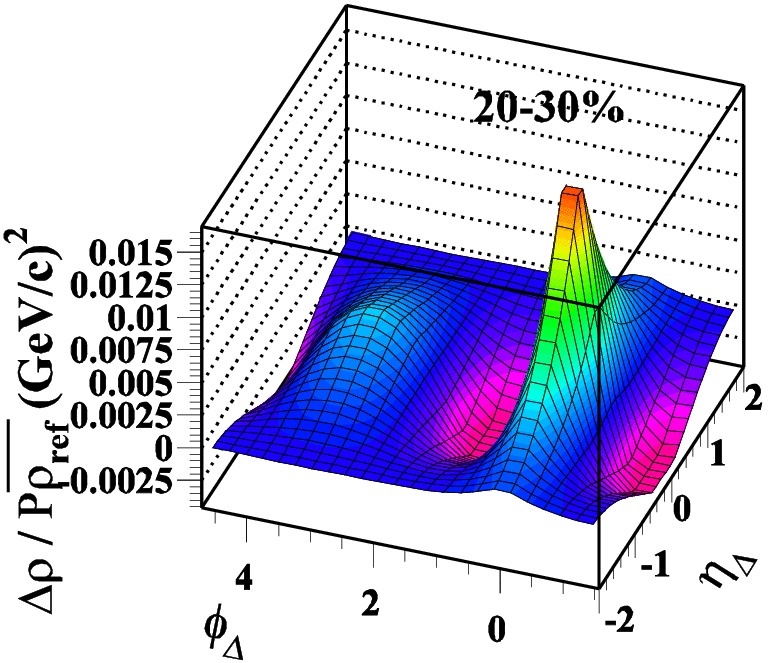}\label{fig11_c}}
    \subfigure[0-5\%]{\includegraphics [width=0.245\textwidth]{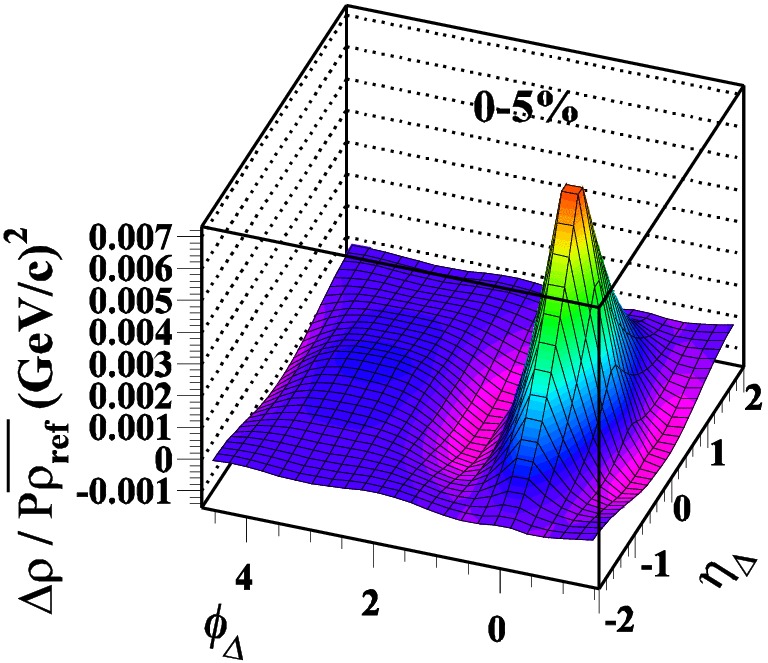}\label{fig11_d}}
    \caption{$p_t$ correlations for 200~GeV AuAu collisions for various centralities.
             These correlations have been inferred from the inversion of scaled fluctuations
             and we have subtracted the multipoles.
             The shapes of the peaks at $(0,0)$ are different than those of
             number correlations and there is a ``recoil'' hole around the peak.
    }
    \label{fig11}
\end{figure}

\section{Summary}

We have presented a detailed differential analysis of minimum-bias jet systematics
in p-p collisions.
We have described low-$Q^2$ jets using fragmentation function systematics.
The widths of the same-side (SS) 2D peak depend on the $y_{t\Sigma}$ of the pair, changing from elongation
along $\phi_\Delta$ to symmetry in $(\eta_\Delta,\phi_\Delta)$, eventually reaching a width described
by $k_t$ scaling as $y_{t\Sigma}$ is increased.
We found that $k_t$ not only affects the away-side (AS) angular correlations but also broadens the
AS $(y_{t1},y_{t2})$ correlations.

For A-A collisions we saw that the SS 2D peak elongates along $\eta_\Delta$, the width increasing
rapidly at a particular centrality (sharp transition).
The amplitude of the SS peak is consistent with binary scaling for peripheral collisions and
greatly exceeds binary-collision scaling starting near the same centrality where the $\eta_\Delta$ width
elongation starts.
The AS dipole amplitude closely follows the SS 2D Gaussian peak amplitude.
In contrast, the $\phi_\Delta$ width of the SS Gaussian decreases slightly with increasing
centrality.
The quadrupole component is small for peripheral and central collisions but has already
become significant before the transition centrality, where particle densities are low.
There is no evidence for an opaque core.
All partons, even low-$Q^2$ partons, are accounted for in the final state.
We are currently studying $p_t$ correlations which provide complementary information on jets.

\end{document}